# Efficient Image Encryption and Decryption Using Discrete Wavelet Transform and Fractional Fourier Transform

*A*

*Dissertation*

*submitted*

*in partial fulfillment*

*for the award of the Degree of*

***Master of Technology***

*in Department of Computer Science and Engineering*

*(with specialization in Computer Science and Engineering)*

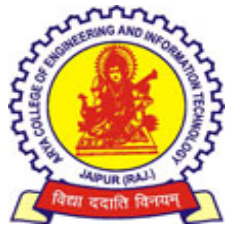

| | |
|---|---|
| Supervisor: | Submitted By: |
| Dr. Shiv Kumar. | Prerana Sharma. |
| Senior Professor. | Enrolment No.: 07E2ARCSE40P61. |

**Department of Computer Engineering**

**Arya College of Engineering and Information Technology, Jaipur**

Rajasthan Technical University

**April, 2013**

I

# Candidate's Declaration

I hereby declare that the work, which is being presented in the Dissertation, entitled **"Efficient Image Encryption and Decryption Using Discrete Wavelet Transform and Fractional Fourier Transform"** in partial fulfillment for the award of Degree of "Master of Technology" in Department of Computer Science and Engineering with Specialization in Computer Science and Engineering**, and submitted to the Department of Computer Science and Engineering**, **Arya College of Engineering and Information Technology, Jaipur, Rajasthan Technical University** is a record of my own investigations carried under the Guidance of Dr. Shiv Kumar, Department of Computer Science and Engineering, Arya College of Engineering and Information Technology, Jaipur**.**

I have not submitted the matter presented in this Dissertation any where for the award of any other Degree.

**(Prerana Sharma)**
Computer Science and Engineering,
Enrolment No.: 07E2ARCSE40P61
Arya College of Engineering and Information Technology,
Jaipur.

**Counter Signed by**
Dr. Shiv Kumar
Department of Computer Science
Arya College of Engineering and Information Technology,
Jaipur.

# ACKNOWLEDGEMENTS

At first, I owe my deepest gratitude to **Dr. Shiv Kumar**, whose guidance, encouragement and positive criticism enabled me to become familiar with the subject. It would have been merely a dream to finish this work without his continuous support and motivation.

I am also thankful to **Mr. Ankur Agarwal, Assistant Professor, Uttarakhand Technical University, Dehradun** for his valuable suggestions on my work. It really helped me a lot to improve the work.

Besides, I would like to express my gratitude to the authors of all the books and publications, which helped me enhancing my knowledge and understanding this domain.

I am indebted to my family for their trust in me that made me confident enough to accomplish this work. Their support in a number of ways must also be acknowledged here.

I would like to thank all my friends for their continuous support throughout my stay at the institute.

Above all, the infinite grace of the Almighty God is of essential importance and I solemnly offer my regards for His grace which enabled peace and harmony for this work.

# CONTENTS

# LIST OF PRINCIPLE SYMBOLS AND ACRONYMS

| | |
|---|---|
| *I* | An image of size *M* X *N* denoted by a 2-*D* function *f*(*x*, *y*) (Sometimes *M*=*N*). |
| $I_1$ | An image of size *M* X *N* denoted by a 2-*D* function *g*(*x*, *y*) (Sometimes *M*=*N*). |
| *I'* | An image of size *M* X *N* denoted by a 2-*D* function *f'*(*x*, *y*) and obtained after decryption of $I_1$ (Sometimes *M*=*N*). |
| *f*(*x*, *y*) | Value of pixel of an image *I*, whose spatial coordinates are *x*, *y*. |
| CRPM | Chaotic Random Phase Mask function. |
| $\phi_1(x, y)$ | Random phase function, which may be chaotic in nature and termed as CRPM in this case. |
| $conj(\phi_1(x, y))$ | Complex conjugate of the random phase function $\phi_1(x, y)$. |
| *f'*(*x*, *y*) | Value of pixel of an image *I'*, whose spatial coordinates are *x*, *y*. |
| *g*(*x*, *y*) | Value of pixel of an image $I_1$, whose spatial coordinates are *x*, *y*. |
| 2-D *FRT* | Two-dimensional fractional Fourier transform. |
| $DWT_2$ | Two-dimensional discrete wavelet transform. |
| 2-D *IFRT* | Two-dimensional inverse fractional Fourier transform. |
| $IDWT_2$ | Two-dimensional inverse discrete wavelet transform. |
| *a* | Order of *FRT*. |
| $F^a$ | $a^{th}$ order *FRT*. |
| *u* | Domain of *FRT*. |
| *u'* | Domain of *FRT*. |
| csc | Trigonometric cosec function. |
| *O*(*N*) | Complexity of the order of *N*. |
| MSE | Mean square error. |

# LIST OF TABLES

# LIST OF FIGURES

# ABSTRACT

Fractional Fourier transform and chaos functions play a key role in many of encryption-decryption algorithms. In this work performance of image encryption-decryption algorithms is quantified and compared using the computation time *i.e.* the time consumption of encryption-decryption process and resemblance of input image to the restored image, quantified by *MSE*.

This work proposes an improvement in computation-time of image encryption-decryption algorithms by utilizing image compression properties of the 2-dimensional Discrete Wavelet Transform ($DWT_2$). Initially, computation complexity of the algorithms is evaluated and compared with that of existing algorithms. This analysis claims the proposed algorithms to be nearly 8 times faster than the existing algorithms.

Further, simulations are performed using *MATLAB*7.7 to quantify performance of existing algorithms and the proposed algorithms using *MSE* and computation time. The results obtained in these simulations prove that for the proposed algorithms *MSE* between restored and original images is lesser than that of existing algorithms thereby maintaining the robustness of the existing algorithms. These algorithms are found sensitive to a variation of $1 \times 10^{-1}$ in the fractional orders used in encryption-decryption process.

# INTRODUCTION

**1.1 Motivation:**

With the rapid development of internet and wide application of multimedia technology, people can communicate the digital multimedia information such as digital image, with others conveniently over the internet. In many cases, image data, transmitted over a network are expected not to be browsed or processed by illegal receivers. Therefore, the security of digital image has attracted much attention recently and many different methods for image encryption have been proposed, such as [1] [2] [3] [4] [5] [6] [7].

Optical systems are of growing interest for image encryption because of their distinct advantages of processing 2-dimensional complex data in parallel at high speed. In the past, many optical methods have been proposed in [1] [2] [4] [6] [8] [9]. Among them the most widely used and highly successful optical encryption scheme is double random phase encoding proposed in [4]. It can be shown that if these random phases are statistically independent white noise then the encrypted image is also a stationary white noise. In some schemes [2] [3] [5] [10], chaos based functions are used to generate random phase mask. As the generalization of the conventional Fourier transform, the fractional Fourier transform has also recently shown its potential in the field of optical security [1] [2] [4] [6] [9].

Image encryption has been very difficult than that of text encryption due to some intrinsic features of images such as bulk data capacity, high correlation among pixels and high redundancy. Fast (Efficient) image encryption has been an area of interest for research due to the need of real-time image encryption-decryption in various fields such as military image transmission etc. [10].

**1.2 Problem Statement:**

This work refers to the algorithms presented by Narendra [2]. Image compression characteristic of $DWT_2$ can be exploited to improve the computation times of existing algorithm which can be verified on the basis of results obtained after executing suitable simulations.

With reference to the three algorithms proposed by Narendra [2], these algorithms shall be improved by inserting $DWT_2$ and $IDWT_2$ in the encryption-decryption process. A mathematical expression for computation complexity of each of these algorithms is to be derived and compared to their existing counterparts thereby producing some theoretical claims which in turn shall be verified.

Simulations are to be performed using existing and proposed algorithms. Performance of these simulations shall be recorded in terms of *MSE* and computation time. Computation time shall be referred as the time required by entire encryption-decryption process.

Besides improving computation time the proposed algorithms are expected to decrease the *MSE* of existing algorithms and maintain the robustness of existing algorithms.

**1.3 Organization of thesis:**

The outline of thesis is as follows. Chapter 2 includes a detailed description of mathematical definition, operational properties and applications of fractional Fourier transform; discrete Fourier transform and wavelet transform. A brief overview of chaos functions and the random phase mask is also included in this chapter.

Three different encryption-decryption algorithms, based on the transforms and functions included in chapter 2 are analyzed in chapter 3. These methods are analyzed on the basis of their quality of encryption-decryption and computation time.

Chapter 4 proposes three new algorithms which are time efficient than the existing algorithms. A mathematical model for encryption and decryption steps of these algorithms is presented in this chapter. Further, there is a detailed discussion about the computation complexity of these algorithms. Section 4.4 includes a theoretical comparison between computation times of the existing and the proposed algorithms.

Chapter 5 summarizes the simulation results obtained using proposed algorithms and their comparison with the existing methods on the basis of various experimental results. Different parameters are defined to quantify the experimental results obtained from the execution of these algorithms. On the basis of these parameters, the theoretical claims made in section 4.4 are verified.

Chapter 6 concludes the comparisons of chapter 5 and suggests future scope of the work done in this thesis.

# LITERATURE SURVEY

## 2.1  Fractional Fourier Transform

Fractional Fourier transform are a one parameter subclass of the class of linear canonical transforms. It is possible to define the fractional Fourier transform in several different ways [11]. Any of these definitions can be taken as a starting point, and the others then derived as properties. Each different definition leads to different physical interpretation which becomes useful in a variety of applications. Following paragraphs introduce some notations and general assumptions.

The $a^{th}$ order fractional Fourier transform of the function $f(u)$ will be denoted in any of the following ways, depending on the context and requirements of clarity. Most commonly, the fractional transform is denoted by $f_a(u)$ or equivalently $F^a f(u)$. The latter expression may be interpreted in two equivalent ways. First, it may be interpreted as the operator $F^a$ acting on the abstract signal $f$. The result of which is expressed in the $u$ domain:

$$f_a(u) \equiv F^a f(u) \equiv (F^a f)(u) \equiv F^a [f](u) \equiv (F^a[f])(u). \qquad (2.1.1)$$

Second, $F^a f(u)$ may be interpreted as the operator $F^a$ acting on the function $f(u)$, with the result again being expressed in the $u$ domain:

$$f_a(u) \equiv F^a f(u) \equiv F^a [f(u)](u) \equiv (F^a [f(u)])(u). \qquad (2.1.2)$$

This second interpretation is appropriate regardless of whether the operator $F^a$ denotes a system or a transformation, whereas the first is appropriate only when $F^a$ denotes a system. Same dummy variable $u$ has been used both for the original function in the time domain, and its fractional Fourier transform. Both explicitly appear in the last two forms in equation (2.1.2), whereas either is absent in the first two.

Further discussion will be restricted to the case where the order parameter '$a$' is a real number. Complex-ordered fractional Fourier transform may be treated as a special case of complex-parameterized linear canonical transforms. It is assumed that $f$ is a finite energy signal. '$u$' is interpreted as a dimensionless variable.

*2.1.1 Definition: Linear integral transform*

The first definition presented here is the most direct and concrete one, although it will not be immediately evident why this transform deserves to be called the fractional Fourier transform. The transform is defined by explicitly specifying its linear transform kernel.

**Definition:** The $a^{th}$ order fractional Fourier transform is a linear operation defined by the integral [11]

$$f_\alpha(u) \equiv \int_{-\infty}^{\infty} K_\alpha(u,u') f(u') du', \tag{2.1.3}$$

$$K_a(u,u') \equiv A_\alpha \exp\left[i\pi\left(\cot\alpha\, u^2 - 2\cosec(\alpha) uu' + \cot\alpha\, u'^2\right)\right], \tag{2.1.4}$$

$$A_\alpha \equiv \sqrt{1 - i\cot\alpha} \qquad \alpha \equiv a\pi/2$$

when $a \neq 2j$ for integer $j$ and $K_a(u, u') = \delta(u - u')$ when $a = 4j$ and $K_a(u, u') = \delta(u + u')$ when $a = 4j\pm2$, where $j$ is an integer. The $a^{th}$ order transform is sometimes referred to as the $\alpha^{th}$ order transform. The square root is defined such that the argument of the result lies in the interval $(-\pi/2, \pi/2]$. For $0 < |a| < 2$ ($0 < |\alpha| < \pi$), $A_\alpha$ can be rewritten without ambiguity as

$$A_\alpha = \exp-i(\pi\,\text{sgn}(\alpha)/4 - \alpha/2)/\sqrt{|\sin\alpha|}, \tag{2.1.5}$$

where sgn (.) is the sign function. When '$a$' is outside the interval $0 \leq |a| \leq 2$, '$a$' is simply replaced by it's modulo 4 equivalent lying in this interval and use this value in equation (2.1.5).

At first sight, this definition does not offer much insight into the nature of the fractional Fourier transform, unless one is very well versed with the class of linear canonical transforms, of which fractional Fourier transforms are easily seen to constitute a one-parameter subclass. Nevertheless equation (2.1.4) is the most direct way of defining the transform. To obtain the $a^{th}$ order fractional Fourier transform of a function $f(u)$, it is simply substituted in the equation (2.1.4).

The transform is by definition linear, but it is not shift-invariant (unless $a = 4j$), since the kernel is not a function of $(u - u')$ only. In order to examine the case where '$a$' is an integer, let $j$ denote an arbitrary integer. It is to be noted that by definition $F^{4j}$ and $F^{4j\pm2}$ correspond to the identity operator $I$ and the parity operator $P$ respectively. For $a = 1$, $\alpha = \pi/2$, $A_\alpha = 1$, and

$$f_1(u) = \int_{-\infty}^{\infty} \exp(-i2\pi u u') f(u') du'. \qquad (2.1.6)$$

Thus $f_1(u)$ is equal to the ordinary Fourier transform of $f(u)$, which until now was denoted as $F(u)$. Likewise, it is possible to see that $f_{-1}(u)$ is the ordinary inverse Fourier transform of $f(u)$. It is further possible to conclude that the above definition of the fractional Fourier transform is consistent with our definition of integer powers of the Fourier transform. Since $\alpha = a\pi/2$ appears in equation (2.1.4) only in the argument of trigonometric functions, the definition is periodic in $a$ with period 4. Thus, further attention will often be limited to the interval $a \in (-2, 2]$ and sometimes $a \in [0, 4)$ (or $\alpha \in [0, 2\pi)$) [11]. These facts can be restated in operator notion:

$$F^0 = I, \qquad (2.1.7)$$
$$F^1 = F, \qquad (2.1.8)$$
$$F^2 = P, \qquad (2.1.9)$$
$$F^3 = FP = PF, \qquad (2.1.10)$$
$$F^4 = F^0 = I, \qquad (2.1.11)$$
$$F^{4j+a} = F^{4j'+a}, \qquad (2.1.12)$$

where $j, j'$ are arbitrary integers.

According to equation (2.1.4), the zero$^{\text{th}}$-order transform of a function is equal to the function it self by definition. Likewise, the $\pm 2^{\text{nd}}$ order transform is equal to $f(-u)$ by definition. This piecewise definition would be rather artificial if it did not exhibit some kind of continuity with respect to '$a$' for all values of '$a$'. It is not difficult to see by examining the kernel that a slight change in $a$ results in only a slight change in $f_a(u)$ when '$a$' is not close to an integer multiple of 2. To see that this is true when '$a$' approaches an integer multiple of 2 as well, first consider the behavior of the kernel as $a \to 0$. For infinitesimal $|a| > 0$ the kernel can be rewritten as

$$K_a(u, u') = \frac{\exp(-i\pi \operatorname{sgn}(\alpha)/4)}{\sqrt{|\alpha|}} \exp[i\pi(u-u')^2/\alpha]. \qquad (2.1.13)$$

This is seen to indeed reduce to $\delta(u-u')$ in the limit $a \to 0$. Alternatively, noting that $K_a(u, u')$ is a function only of $(u-u')$, the function can be defined as

$$K_a(u) \equiv \frac{\exp(-i\pi \operatorname{sgn}(\alpha)/4)}{\sqrt{|\alpha|}} \exp[i\pi(u)^2/\alpha], \qquad (2.1.14)$$

which is convolved with $f(u)$ to obtain $K_a(u) * f(u) = f_a(u)$. The Fourier transform of $K_a(u)$, as $\exp(-i\pi\alpha\mu^2)$, approaches unity as $\alpha \to 0$, which in turn implies that $K_a(u)$ approaches a delta function. Thus, the definition of the transform is indeed continuous with respect to $a=0$. A similar discussion is possible when $a$ approaches other integer multiples of 2.

Another, very important property of the fractional Fourier transform operator, the index additively property, can be stated in the alternative forms

$$F^{a_1}[F^{a_2}[f]] = F^{a_1+a_2}[f] = F^{a_2}[F^{a_1}[f]], \qquad (2.1.15)$$

This can be proved by repeated application of equation (2.1.4), a process which is complicated by the square root appearing in the coefficient $A_\alpha$. This process amounts to showing

$$\int K_{a_2}(u,u'') K_{a_1}(u'',u') du'' = K_{a_1+a_2}(u,u') \qquad (2.1.16)$$

by direct integration, which can be accomplished by using Gaussian integrals. Its proof may be found in [12].

Since Fractional Fourier transform are linear canonical transforms, they also satisfy the associative property, as well as other properties of linear canonical transforms. In particular, fractional Fourier transforms are unitary, as the kernel of the inverse transform can be obtained by replacing '$a$' with '$-a$':

$$K_a^{-1}(u,u') = K_{-a}(u,u') = K_a^*(u,u') = K_a^*(u',u) = K_a^H(u,u'). \qquad (2.1.17)$$

Note that the kernel $K_a(u, u')$ is symmetric, but not Hermitian. Unitarily implies that the fractional Fourier transform can be interpreted as a transformation from one representation to another, and that inner products and norms are not changed under the transform.

Now, the first interpretation of the fractional Fourier transform is given as follows. Let us concentrate on the interval $0 \leq a \leq 1$. It is known that when $a = 0$ the fractional Fourier transform is the original function and when $a = 1$ it is the ordinary Fourier transform. As '$a$' varies from 0 to 1, the transform evolves smoothly from the original function to the ordinary

Fourier transform. The fact that the fractional Fourier transform interpolates between the original function and its ordinary Fourier transform with the continuous parameter *a,* offers some justifications for its name.

As a final comment, a transform can be interested as a system mapping "input" functions $f(u)$ to "output" functions $f_a(u)$. In this interpretation, the same dummy variable $u$ is used for both the input and outputs. This notation is usually found to be the most convenient. On the other hand, the functions $f_a(.)$ for different values of '*a*' may be interpreted as different representations of the same abstract signal $f$ and $f_a(.)$ may be considered as being the representation of the signal $f$ in the $a^{th}$ order *fractional Fourier domain*. In this case it is often useful to distinguish the variables associated with each domain by labeling them as $u_a$. Thus $f_a(u_a)$ is the representation in the $a^{th}$ domain, $f_0(u_0)$ is the representation in the time domain, and $f_1(u_1)$ is the frequency-domain representation. The axis $u_a$ may be referred to as the $a^{th}$ fractional Fourier domain, so that $u_0$ and $u_1$ are the conventional time and frequency domains $u$ and $\mu$. The representation of the signal in the $a^{th}$ domain can be obtained from its representation in the $a^{th}$ domain through an $(a' - a)^{th}$ order fractional Fourier transformation:

$$f_{a'}(u_{a'}) = \int K_{a'-a}(u_{a'}, u_a) f_a(u_a) du_a .$$  (2.1.18)

*2.1.2 Properties of fractional Fourier transform*

**TABLE 2.1**
**PROPERTIES OF FRACTIONAL FOURIER TRANSFORM [11]**

| | |
|---|---|
| 1. Linearity | $F^a[\sum_j \alpha_j f_j(u)] = \sum_j \alpha_j [F^a f_j(u)]$ |
| 2. Integer orders | $F^j = (F)^j$ |
| 3. Inverse | $(F^a)^{-1} = F^{-a}$ |
| 4. Unitarity | $(F^a)^{-1} = (F^a)^H$ |
| 5. Index additivity | $F^{a_2} F^{a_1} = F^{a_2+a_1}$ |
| 6. Commutativity | $F^{a_2} F^{a_1} = F^{a_1} F^{a_2}$ |
| 7. Associativity | $F^{a_3}(F^{a_2} F^{a_1}) = (F^{a_3} F^{a_2}) F^{a_1}$ |
| 8. Eignfunctions | $F^a \psi_l = \exp(-ial\pi/2) \psi_l$ |
| 9. Wigner distribution | $W_{f_a}(u, \mu) = W_f(u\cos\alpha - \mu\sin\alpha, u\sin\alpha + \mu\cos\alpha)$ |
| 10. Parseval | $\langle f(u), g(u) \rangle = \langle f_a(u), g_a(u) \rangle$ |

*2.1.3 Operational Properties of fractional Fourier transform*

**TABLE 2.2**
**OPERATIONAL PROPERTIES OF FRACTIONAL FOURIER TRANSFORM [11]**

| $f(u)$ | $f_a(u)$ |
|---|---|
| 1. $f(-u)$ | $f_a(-u)$ |
| 2. $|M|^{-1} f(u/M)$ | $\sqrt{\dfrac{1 - i \cot \alpha}{1 - iM^2 \cot \alpha}} \exp\left[i\pi u^2 \cot \alpha \left(1 - \dfrac{\cos^2 \alpha'}{\cos^2 \alpha}\right)\right] f_{a'}\left(\dfrac{Mu \sin \alpha'}{\sin \alpha}\right)$ |
| 3. $f(u - \xi)$ | $\exp(i\pi \xi^2 \sin \alpha \cos \alpha) \exp(-i2\pi u \xi \sin \alpha) f_a(u - \xi \cos \alpha)$ |
| 4. $\exp(i2\pi \xi u) f(u)$ | $\exp(-i\pi \xi^2 \sin \alpha \cos \alpha) \exp(i2\pi u \xi \cos \alpha) f_a(u - \xi \sin \alpha)$ |
| 5. $u^n f(u)$ | $\left[\cos \alpha u - \sin \alpha (i2\pi)^{-1} d/du\right]^n f_a(u)$ |
| 6. $\left[(i2\pi)^{-1} d/du\right]^n f(u)$ | $\left[\sin \alpha u + \cos \alpha (i2\pi)^{-1} d/du\right]^n f_a(u)$ |
| 7. $f(u)/u$ | $-i \csc \alpha \exp(i\pi u^2 \cot \alpha) \int_{-\infty}^{2\pi u} f_a(u') \exp(-i\pi u'^2 \cot \alpha) du'$ |
| 8. $\int_\xi^u f(u') du'$ | $\sec \alpha \exp(-i\pi u^2 \tan \alpha) \int_\xi^u f_a(u') \exp(i\pi u'^2 \tan \alpha) du'$ |
| 9. $f^*(u)$ | $f_{-a}^*(u)$ |
| 10. $f^*(-u)$ | $f_{-a}^*(-u)$ |
| 11. $[f(u) + f(-u)]/2$ | $[f_a(u) + f_a(-u)]/2$ |
| 12. $[f(u) - f(-u)]/2$ | $[f_a(u) - f_a(-u)]/2$ |

*2.1.4 Fractional Fourier transform of some common functions*

**TABLE 2.3**
**FRACTIONAL FOURIER TRANSFORM OF SOME COMMON FUNCTIONS [11]**

| | |
|---|---|
| 1. $\delta(u)$ | $\sqrt{1 - i \cot \alpha} \exp(i\pi u^2 \cot \alpha)$ |
| 2. $\delta(u - \xi)$ | $\sqrt{1 - i \cot \alpha} \exp[i\pi(u^2 \cot \alpha - 2u\xi \csc \alpha + \xi^2 \cot \alpha)]$ |
| 3. $1$ | $\sqrt{1 - i \tan \alpha} \exp(-i\pi u^2 \tan \alpha)$ |
| 4. $\exp(i2\pi \xi u)$ | $\sqrt{1 + i \tan \alpha} \exp[-i\pi(u^2 \tan \alpha - 2u\xi \sec \alpha + \xi^2 \tan \alpha)]$ |
| 5. $\exp(i\pi \chi u^2)$ | $\sqrt{\dfrac{1 + i \tan \alpha}{1 + \chi \tan \alpha}} \exp\left[i\pi u^2 \dfrac{\chi - \tan \alpha}{1 + \chi \tan \alpha}\right]$ |

6. $\exp[i\pi(\chi u^2 + 2\chi u)]$  $\sqrt{\dfrac{1+i\tan\alpha}{1+\chi\tan\alpha}} \exp\left[i\pi \dfrac{u^2(\chi - \tan\alpha) + 2u\xi\sec\alpha - \xi^2 \tan\alpha}{1+\chi\tan\alpha}\right]$

7. $\psi_l(u)$  $\exp(-il\alpha)\psi_l(u)$

8. $\exp(-\pi u^2)$  $\exp(-\pi u^2)$

9. $\exp(-\pi\chi u^2)$  $\sqrt{\dfrac{1-i\cot\alpha}{\chi - i\cot\alpha}} \exp\left[i\pi u^2 \dfrac{\cot\alpha(\chi^2 - 1)}{\chi^2 + \cot^2\alpha}\right] \exp\left[-\pi u^2 \dfrac{\chi\csc^2\alpha}{\chi^2 + \cot^2\alpha}\right]$

10. $\exp[-\pi(\chi u^2 + 2\xi u)]$  $\sqrt{\dfrac{1-i\cot\alpha}{\chi - i\cot\alpha}} \exp\left[i\pi\cot\alpha \dfrac{u^2(\chi^2 - 1) + 2u\xi\sec\alpha + \xi^2}{\chi^2 + \cot\alpha}\right] \times$
$\exp\left[-\pi\csc^2\alpha \dfrac{u^2\chi + 2u\chi\cos\alpha - \chi\xi^2\sin^2\alpha}{\chi^2 + \cot\alpha}\right]$

---

## *2.1.5 Applications of Fractional Fourier transform*

- Fractional Fourier Transform is used for Filtering, Estimation, and Signal Recovery.
- Fractional Fourier Transform is used for Matched Filtering, Detection, and Pattern Recognition
- Fractional Fourier Transform is also used in Optics.

## 2.2 Discrete Fourier Transform.

### *2.2.1 Definitions of the Discrete Fourier Transform*

The discrete Fourier transform pair is defined by [11]

$$X(k) \equiv F_D\{x(n)\} = \sum_{n=0}^{N-1} x(n) e^{-j2\pi kn/N}, k = 0,1,2,\ldots,N-1 \qquad (2.2.1)$$

$$x(n) \equiv F_D^{-1}\{X(k)\} = \frac{1}{N}\sum_{k=0}^{N-1} X(k) e^{j2\pi kn/N}, n = 0,1,2,\ldots,N-1 \qquad (2.2.2)$$

where $X(k) \equiv X(2\pi k/N)$. Substituting equation (2.2.1) in equation (2.2.2), gives

$$\frac{1}{N}\sum_{k=0}^{N-1}\left[\sum_{m=0}^{N-1} x(m) e^{-j2\pi km/N}\right] e^{j2\pi km/N} = \frac{1}{N}\sum_{m=0}^{N-1} x(m) \sum_{k=0}^{N-1} e^{-j2\pi k(m-n)/N}$$

But the last summation is equal to zero for $m \neq n$ and equal to one for $m=n$ and, thus, the last expression becomes $x(n)N/N = x(n)$ which proves that equations (2.2.1) and (2.2.2) are the DFT pair.

### *2.2.1.1 DFT as a Linear Transformation*

Let $x_N$ be an N-point vector of the signal sequence and $X_N$ is an N-point sequence of the frequency samples. Equation (2.2.1) can be written in the form

$$\underline{X}_N = \underline{W}_N \underline{x}_N \qquad (2.2.3)$$

where

$$X_N = \begin{bmatrix} X(0) \\ X(1) \\ \vdots \\ X(N-1) \end{bmatrix}, \quad x_N = \begin{bmatrix} x(0) \\ x(1) \\ \vdots \\ x(N-1) \end{bmatrix},$$

$$W_N = \begin{bmatrix} 1 & 1 & 1 & \cdots & 1 \\ 1 & e^{-j2\pi/N} & (e^{-j2\pi/N})^2 & & (e^{-j2\pi/N})^{N-1} \\ \vdots & & & & \vdots \\ 1 & (e^{-j2\pi/N})^{N-1} & (e^{-j2\pi/N})^{N-1} & \cdots (e^{-j2\pi/N})^{(N-1)(N-1)} \end{bmatrix} \qquad (2.2.4)$$

If the inverse of $\underline{W}_N$ exists, then Equation (2.2.3) gives

$$x_N = W_N^{-1} X_N \qquad (2.2.5)$$

which is the inverse Discrete Fourier transform (IDFT). The matrix $\underline{W}_N$ is symmetric has the properties

$$\underline{W}_N^{-1} = \frac{1}{N} \underline{W}_N^*, \quad W_N W_N^* = N \underline{I}_N \qquad (2.2.6)$$

Where, $I_N$ is an $N \times N$ identity matrix

$\underline{W}_N$ is an orthogonal (Unitary matrix)

### *2.2.2 Properties of the DFT*

**TABLE 2.4**
**PROPERTIES OF THE DFT [11]**

| Property | Time functions $x(n), h(n)$ | Frequency domain Functions $X(k), H(k)$ |
|---|---|---|
| Linearity | $ax(n) + bh(n)$ | $aX(k) + bH(k)$ |
| Periodicity | $x(n) = x(n + N)$ | $X(k) = X(k + N)$ |
| Time reversal | $x(N - n)$ | $X(N - k)$ |
| Circular time shift | $x((n - \ell))_N$ | $X(k)e^{-j2\pi k\ell/N}$ |
| Circular frequency shift | $x(n)e^{j2\pi \ell n/N}$ | $X((k - \ell))_N$ |
| Complex conjugate | $x^*(n)$ | $X^*(N - k)$ |
| Circular convolution | $x(n) \otimes h(n)$ | $X(k) H(k)$ |
| Circular correlation | $x(n) \otimes h^*(-n)$ | $X(k) H^*(k)$ |
| Multiplication | $x(n)h(n)$ | $\frac{1}{N} X(k) \otimes H^*(k)$ |
| Symmetry | $\frac{1}{N} X(n)$ | $x(-k)$ |
| Parseval's theorem | $\sum_{n=0}^{N-1} x(n)h^*(n) = \frac{1}{N} \sum_{n=0}^{N-1} X(k) H^*(k)$ | |

### *2.2.3 DFT of various functions*

**TABLE 2.5**
**DFT OF VARIOUS FUNCTIONS [11]**

| *f(n)* | *F(k)* |
|---|---|
| $\delta(n - n_0), \ 0 \leq n_0 \leq N-1,$ $n_0 = $ integer, $0 \leq n \leq N-1$ | $\exp(-j2\pi n_0 k / N), \ 0 \leq k \leq N-1$ |
| $\exp(j2\pi k_0 / N), \ 0 \leq k_0 \leq N-1,$ $k_0 = $ integer, $0 \leq n \leq N-1$ | $\delta(k - k_0), \ 0 \leq k \leq N-1$ |
| $u(n) - u(n - N), 0 \leq n \leq N-1$ | $F(k) = \begin{cases} 1, & k = 0 \\ 0, & \text{otherwise} \end{cases}$ |
| $\cos(2\pi k_0 n / N), 0 \leq k_0 \leq N-1,$ $k_0 = $ integer, $0 \leq n \leq N-1$ | $F(k) = \begin{cases} \frac{N}{2} \delta(k - k_0) & k = k_0 \\ \frac{N}{2} \delta[k - (N - k_0)] & k = N - k_0 \end{cases}$ |
| $\cos(\pi n), \ 0 \leq n \leq N-1$ | $F(k) = N\delta\left(k - \frac{N}{2}\right), \ 0 \leq k \leq N-1$ |
| $\cos\left(\frac{\pi n}{N}\right), \ 0 \leq n \leq N-1$ | $F(k) = \frac{1}{2} \frac{1 - \exp[-j(2\pi k - \pi)]}{1 - \exp\left[-j\left(\frac{2\pi k}{N} - \frac{\pi}{N}\right)\right]}$ $+ \frac{1}{2} \frac{1 - \exp[-j(2\pi k + \pi)]}{1 - \exp\left[-j\left(\frac{2\pi k}{N} + \frac{\pi}{N}\right)\right]}, 0 \leq k \leq N-1$ |
| $n/N, \quad 0 \leq n \leq N-1$ | $F(k) = \begin{cases} \dfrac{N}{2} & k = 0 \\ j\dfrac{\cos(k\pi)\sin(k\pi/N)}{2\sin^2(k\pi/N)} & 0 \leq k \leq N-1 \end{cases}$ |
| $\begin{cases} 1 & 0 \leq n \leq m \\ 0 & m < n \leq N-1 \end{cases}$ | $F(k) = \exp\left(-j\dfrac{\pi km}{N}\right)\dfrac{\sin\left(\dfrac{\pi k(m+1)}{N}\right)}{\sin\dfrac{\pi k}{N}}$, $0 \leq k \leq N-1$ |

## 2.3 Wavelet transform (HAAR)

The wavelet transform is a new mathematical tool developed mainly since the middle of the 1980s. It is efficient for local analysis of nonstationary and fast transient wideband signals. The wavelet transform is a mapping of a time signal to the timescale joint representation, which is used in the short-time Fourier transform, the Wigner distribution and the ambiguity function. The temporal aspect of the signals is preserved. The wavelet transform provides multiresolution analysis with dilated windows. The higher frequency analysis is done using narrower windows and the lower frequency analysis is done using wider windows. Thus, the wavelet transform is a constant-$Q$ analysis. The basis functions of the wavelet transform, the wavelets, are generated from a basic wavelet function by dilations and translations. They satisfy an admissible condition so that the original signal can be reconstructed by the inverse wavelet transform. The wavelets satisfy also the regularity condition so that the wavelet coefficients decrease fast with the decreasing of the scale. The wavelet transform is local not only in time but also in frequency domain.

To reduce the time–bandwidth product of the wavelet transform output, the discrete wavelet transform with discrete dilations and translations of the continuous wavelets can be used. The orthonormal wavelet transform is implemented in the multiresolution signal analysis framework, which is based on the scaling functions. The discrete translates of the scaling functions form an orthonormal basis at each resolution level. The wavelet basis is generated from the scaling function basis. The two bases are mutually orthogonal at each resolution level. The scaling function is an averaging function. The orthogonal projection of a function onto the scaling function basis is an averaged approximation. The orthogonal projection onto the wavelet basis is the difference between two approximations at two adjacent resolution levels. Both the scaling functions and the wavelets satisfy the orthonormality conditions and the regularity conditions. The discrete orthonormal wavelet series decomposition and reconstruction are computed in the multiresolution analysis framework with recurring two discrete low-pass and high-pass filters, that are, in fact, the 2-band paraunitary perfect reconstruction quadrature mirror filters, developed in the subband coding theory, with the additional regularity. The tree algorithm operating the discrete wavelet transform requires only $O(L)$ operations where $L$ is the length of the data vector [11]. The time–bandwidth product of the wavelet transform output is only slightly increased with respect to that of the signal.

The wavelet transform is powerful tool for multiresolution local spectrum analysis of nonstationary signals, such as the sound, radar, sonar, seismic, electrocardiographic signals, and for image compression, image processing and pattern recognition. In this chapter all integrations extend from -1 to 1, unless stated otherwise. The wavelet transform can be easily generalized to any dimensions.

*2.3.1 Overview of Wavelet Transform.*

*2.3.1.1 Continuous Wavelet Transform* **[11]**

**Definition** Let $L^2(R)$ denote the vector space of measurable, square-integrable functions. The continuous wavelet transform of a function $f(t) \in L^2(R)$ is a decomposition of $f(t)$ into a set of basis functions $h_{s,\tau}(t)$ called the wavelets:

$$W_f(s,\tau) = \int f(t) h_{s,\tau}^*(t) dt \qquad (2.3.1)$$

where * denotes the complex conjugate. However, most wavelets are real valued. The wavelets are generated from a single basic wavelet (mother wavelet) $h(t)$ by scaling and translation:

$$h_{s,\tau}(t) = \frac{1}{\sqrt{s}} h\left(\frac{t-\tau}{s}\right) \qquad (2.3.2)$$

where, $s$ is the scale factor

$\tau$ is the translation factor.

Usually scale factor $s>0$ is considered to be only positive. The wavelets are dilated when the scale $s>1$ and are contracted when $s<1$. The wavelets $h_{s,\tau}(t)$ generated from the same basic wavelet have different scales $s$ and locations $t$, but all have the identical shape.

The constant $s^{-1/2}$ in the equation (2.3.2) of the wavelets is for energy normalization. The wavelets are normalized in terms of energy as:

$$\int |h_{s,\tau}(t)|^2 dt = \int |h(t)|^2 dt = 1 \qquad (2.3.3)$$

so that all the wavelets scaled by the factor $s$ would have the same energy. The wavelets can also be normalized in terms of amplitude as

$$\int |h_{s,\tau}(t)| dt = 1 \qquad (2.3.4)$$

In this case, the normalization constant is $s^{-1}$ instead of $s^{-1/2}$, and the wavelets are generated from the basic wavelet as

$$h_{s,\tau}(t) = \frac{1}{s} h\left(\frac{t-\tau}{s}\right) \qquad (2.3.5)$$

On substituting equation (2.3.2) into equation (2.3.1) the wavelet transform of $f(t)$ is expressed as a correlation between the signal and the scaled wavelets $h(t/s)$:

$$W_f(s,\tau) = \frac{1}{\sqrt{s}} \int f(t) h^*\left(\frac{t-\tau}{s}\right) dt \qquad (2.3.6)$$

*2.3.1.1.1 Wavelet Transform in Frequency Domain*

The Fourier transform of the wavelet is

$$H_{s,\tau}(\omega) = \int \frac{1}{\sqrt{s}} h\left(\frac{t-\tau}{s}\right) \exp(-j\omega t) dt$$
$$= \sqrt{s} H(s\omega) \exp(-j\omega \tau) dt \qquad (2.3.7)$$

where $H(\omega)$ is the Fourier transform of the basic wavelet $h(t)$. In the frequency domain the Fourier transform of the wavelet is scaled by $1/s$, multiplied by a phase factor $\exp(-j\omega\tau)$ and by a normalization factor $s^{1/2}$. The amplitude of the scaled wavelet is proportional to $s^{-1/2}$ in the time domain and is proportional to $s^{1/2}$ in the frequency domain. Note that when the wavelets are normalized in terms of amplitude, their Fourier transforms of different scales will have the same amplitude. This is suitable for implementation of the continuous wavelet transform using the frequency domain filtering.

Equation (2.3.7) shows a well known concept that a dilation $t/s$ ($s>1$) of a function in the time domain produces a contraction $s\omega$ of its Fourier transform. The term $1/s$ has a dimension of frequency and is equivalent here to the frequency. However, the term ''scale'' is preferred to the term ''frequency'' for the wavelet transform. The term ''frequency'' is reserved to be a parameter related to the Fourier transform.

The correlation between the signal and the wavelets in the time domain can be written as the inverse Fourier transform of the product of the Fourier transform of the wavelets and the Fourier transform of the signal:

$$W_f(s,\tau) = \frac{\sqrt{s}}{2\pi} \int F(\omega) H^*(s\omega) \exp(j\omega\tau) d\omega \qquad (2.3.8)$$

The Fourier transforms of the wavelets $\sqrt{s}H(s\omega)$ are referred to as the wavelet transform filters, and the impulse response of the wavelet transform filter is the scaled wavelet $s^{-1/2}h(t/s)$, where the explicit phase shift exp ($j\omega\tau$) in the frequency and translation $\tau$ in the time are removed. Therefore, the wavelet transform is a bank of wavelet transform filters with different scales $s$.

In the definition of the wavelet transform, the kernel function, wavelet, is not specified. This is a difference between the wavelet transform and other transforms such as the Fourier transform. The theory of wavelet transform deals with general properties of the wavelet and the wavelet transform, such as the admissibility, regularity, and orthogonality. The wavelet basis is built to satisfy these basic conditions. The wavelets can be given as analytical or numerical functions. They can be orthonormal or nonorthonormal, continuous or discrete. One can choose or even builds a proper wavelet basis for a specific application. Therefore, when talking about the wavelet transform one used to specify what wavelet is used in the transform.

The most important properties of the wavelets are the admissibility and regularity. According to the admissible condition the wavelet must oscillate to have its mean value equal to zero. According to the regularity condition the wavelet have exponential decay so that its first low order moments are equal to zero. Therefore, in the time domain the wavelet is just like a small wave that oscillates and vanishes, as that described by the name wavelet. The wavelet transform is a local operator in the time domain. The orthonormality is a property which belongs to the discrete wavelet transform.

**2.3.2 Properties of the Wavelets**

This section discusses some basic properties of the wavelets. One of them is related to the fact that one must be able to reconstruct the signal from its wavelet transform. This property involves the resolution of identity, the energy conservation in the timescale space and the wavelet admissible condition. First any square integrable function which has finite energy and satisfies the wavelets admissible condition can be a wavelet. The second basic property is related to the fact that the wavelet transform should be a local operator in both time and frequency domains. Hence, the regularity condition is usually imposed on the wavelets. The third basic property is related to the fact that the wavelet transform is a multiresolution signal analysis.

*2.3.2.1 Admissible Condition*

*2.3.2.1.1 Resolution of Identity*

The wavelet transform of a 1-D signal is a 2-D timescale joint representation. No information should be lost during the wavelet transform. Hence, the resolution of identity must be satisfied, that is expressed as

$$\int \frac{ds}{s^2} \int d\tau \langle f_1, h_{s,\tau} \rangle \langle h_{s,\tau}, f_2 \rangle = c_h \langle f_1 f_2 \rangle \qquad (2.3.9)$$

where $\langle \ \rangle$ denotes the inner product so that $\langle f_1, h_{s,\tau} \rangle$ is the wavelet transform of $f_1$ as defined in equation (2.3.6) and $c_h$ is a constant. In the left-hand side of equation (2.3.9) the extra factor $1/s^2$ in the integral is the Haar invariant measure, owing to the timescale space differential elements, $d\tau d(1/s) = d\tau ds/s^2$. Here, positive dilation i.e. $s>0$ is assumed. Using the expression for the wavelet transform in the Fourier domain

$$\int \frac{ds}{s^2} \int d\tau < f_1, h_{s,\tau} >< h_{s,\tau}, f_2 >$$

$$= \frac{1}{4\pi^2} \int \frac{ds}{s^2} \int d\tau \iint s F_1(\omega_1) H^*(s\omega_1) F_2^*(\omega_2) H(s\omega_2) e^{j\tau(\omega_1 \omega_2)} d\omega_1 d\omega_2$$

$$= \frac{1}{2} \iint F_1(\omega_1) F_2^*(\omega_1) |H(s\omega_1)|^2 \frac{ds}{s} d\omega_1$$

$$= \frac{c_h}{2\pi} \iiint F_1(\omega_1) F_2^*(\omega_2) d\omega_1 \qquad (2.3.10)$$

where by using change of variables $\omega = s\omega_1$ and $ds = d\omega/|\omega_1|$, so that $ds$ and $d\omega$ are of the same sign. Because $s > 0$, $ds/s = d\omega/|\omega|$, then

$$c_h = \int |H(\omega^2)| \frac{d\omega}{|\omega|} \qquad (2.3.11)$$

According to the Parseval's equality in the Fourier transform we have

$$\frac{1}{2\omega}\int F_1(\omega_1)F_2^*(\omega_1)d\omega_1 = \int f_1 f_2^*(t)dt = <f_1, f_2> \quad (2.3.12)$$

Hence, the resolution of identity is satisfied on the condition that

$$c_h = \int \frac{|H(\omega)|^2}{|\omega|}d\omega < +\infty \quad (2.3.13)$$

### 2.3.2.1.2 Admissible Condition

The condition in equation (2.3.13) is the admissible condition of the wavelet, which implies that the Fourier transform of the wavelet must be equal to zero at the zero frequency:

$$|H(\omega)|^2\big|_{\omega=0} = 0 \quad (2.3.14)$$

Equivalently, in the time domain the wavelet must be oscillatory, like a wave, to have a zero-integrated area, or a zero-mean value:

$$\int h(t)dt = 0 \quad (2.3.15)$$

### 2.3.2.1.3 Energy Conservation

When $f_1 = f_2$, the resolution of identity, equation (2.3.6) becomes

$$\iint |W_f(s,\tau)|^2 d\tau \frac{ds}{s^2} = c_h \int |f(t)|^2 dt \quad (2.3.16)$$

This is the energy conservation relation of the wavelet transform, equivalent to the Parseval energy relation in the Fourier transform.

### 2.3.2.1.4 Inverse Wavelet Transform

From equation (2.3.6) the inverse wavelet transform is obtained as

$$f(t) = \frac{1}{c_h}\iint W_f(s,\tau)\frac{1}{\sqrt{s}}h\left(\frac{t-\tau}{s}\right)d\tau \frac{ds}{s^2} \quad (2.3.17)$$

The function $f(t)$ is recovered from the inverse wavelet transform by the integrating in the timescale space the wavelets $h_{s,\tau}(t)$ weighted by the wavelet transform coefficients $W_f(s, \tau)$.

Thus, the wavelet transform is a decomposition of a function into a linear combination of the wavelets. The wavelet transform coefficients $W_f(s, \tau)$ are the inner products between the function and the wavelets. The $W_f(s, \tau)$ indicate how close the function $f(t)$ is to the corresponding basis functions $h_{s,\tau}(t)$.

### 2.3.2.1.5 Regularity of Wavelet

For the sake of simplicity, consider translation of the wavelet for $\tau = 0$ and the convergence to zero of the wavelet transform coefficients with increasing $1/s$ and decreasing *s*. The signal *f(t)* is expanded into the Taylor series at *t*=0 until order *n*. The wavelet transform coefficients become [13]

$$W_f(s,0) = \frac{1}{\sqrt{s}} \int f(t) h^*\left(\frac{t}{s}\right) dt$$

$$= \frac{1}{\sqrt{s}} \left[ \sum_{p=0}^{n} f^{(p)}(0) \int \frac{t^p}{p!} h\left(\frac{t}{s}\right) dt + \int R(t) h\left(\frac{t}{s}\right) dt \right] \quad (2.3.18)$$

where the remainder in the Taylor series is

$$R(t) = \int_0^t \frac{(t-t')^n}{n!} f^{(n+1)}(t') dt'$$

and $f^{(p)}(0)$ denotes the $p^{th}$ derivative of *f(t)* at *t*=0. Denoting the moments of the wavelets by $M_p$

$$M_p = \int t^p h(t) dt \quad (2.3.19)$$

it is easy to show that the last term in the right-hand side of equation (2.3.18) which is the wavelet transform of the remainder, decreases as $s^{n+2}$. Then a finite development is obtained as

$$W_f(s,0) = \frac{1}{\sqrt{s}} \left[ f(0) M_0 s + \frac{f'(0)}{1!} M_1 s^2 + f''(0) M_2 s^3 + \cdots \right.$$

$$\left. \cdots + \frac{f^{(n)}(0)}{n!} M_n s^{n+1} + O(s^{n+2}) \right] \quad (2.3.20)$$

According to the admissible condition of the wavelet, $M_0$=0, the first term in the right-hand side of equation (2.3.20) must be zero. The speed of convergence to zero of the wavelet transform coefficients $W_f(s, t)$ with decreasing of the scale *s* or increasing of 1/*s* is then determined by the first nonzero moment of the basic wavelet *h(t)*. It is in general required that the wavelets have the first *n*+1 moments until order *n*, equal to zero:

$$M_p = \int t^p h(t) dt = 0 \quad \text{for } p = 0,1,2,\ldots,n \quad (2.3.21)$$

Then, according to equation (2.3.20) the wavelet transform coefficient $W_f(s, t)$ decays as fast as $s^{n+(1/2)}$ for a smooth signal *f(t)*. This regularity condition leads to localization of the wavelet transform in the frequency domain. The wavelet satisfying the condition (Equation 2.3.21) is called the wavelet of order *n*. In frequency domain, this condition is equivalent to the derivatives of the Fourier transform of the wavelet *h(t)* up to order *n* to be zero at the zero frequency ω = 0.

*2.3.2.3 Multi-resolution Wavelet Analysis*

The wavelet transform performs the multi-resolution signal analysis with the varying scale factor *s*. The purpose of the multi-resolution signal analysis is decomposing the signal in multiple frequency bands in order to process the signal in multiple frequency bands differently and independently. Hence, the wavelet needs to be local in both time and frequency domains. Historically, looking for a kernel function which is local in both time and frequency domains has been a hard research topic and has led to invention of the wavelet transform.

*2.3.2.3.1 Localization in Time Domain*

According to the admissible condition the wavelet must oscillate to have a zero mean. According to the regularity condition the wavelet of order *n* has first *n*+1 vanishing moments and decays as fast as $t^{-n}$. Therefore, in the time domain the wavelet must be a small wave that oscillates and vanishes, as that described by the name wavelet. The wavelet is localized in the time domain.

*2.3.2.3.2 Localization in Frequency Domain*

According to the regularity condition the wavelet transform with a wavelet of order *n* decays with *s* as $s^{n+(1/2)}$ for a smooth signal. According to the frequency domain wavelet transform, when the scale *s* decreases the wavelet $H(s\omega)$ in the frequency domain is dilated to cover a large frequency band of the signal Fourier spectrum. Therefore, the decay with *s* as $s^{n+(1/2)}$ of the wavelet transform coefficient implies that the Fourier transform of the wavelet must decay fast with the frequency $\omega$. The wavelet must be local in frequency domain.

*2.3.2.3.3 Band-Pass Filters*

In the frequency domain, the wavelet is localized according to the regularity condition, and is equal to zero at the zero frequency according to the admissible condition. Therefore, the wavelet is intrinsically a band-pass filter.

*2.3.2.4 Linear Transform Property*

By definition the wavelet transform is a linear operation. Given a function *f(t)*, its wavelet transform $W_f(s, \tau)$ satisfies the following relations

1. Linear superposition without the cross terms

$$W_{f_1+f_2}(s,\tau) = W_{f_1}(s,\tau) + W_{f_2}(s,\tau), \qquad (2.3.22)$$

2. Translation

$$W_{f(t-t_0)}(s,\tau) = W_{f(t)}(s,\tau - t_0) \qquad (2.3.23)$$

3. Rescale

$$W_{\alpha^{1/2} f(\alpha t)}(s,\tau) = W_{f(t)}(\alpha s, \alpha \tau) \qquad (2.3.24)$$

Different from the standard Fourier transform and other transforms, the wavelet transform is not ready for closed form solution apart from some very simple functions such as:

**1.** For $f(t)=1$, from the definition (Equation 2.3.6) and the admissible condition of the wavelets, Equation (2.3.15)

$$W_f(s,\tau) = 0 \qquad (2.3.25)$$

The wavelet transform of a constant is equal to zero.

**2.** For a sinusoidal function $f(t)=\exp(j\omega_0 t)$, directly from the Fourier transform of the wavelets

$$W_f(s,\tau) = \sqrt{s} H^*(s\omega_0) \exp(j\omega_0 \tau) \qquad (2.3.26)$$

The wavelet transform of a sinusoidal function is a sinusoidal function of the time shift $\tau$. Its modulus $|W_f(s,\tau)|$ depends on the scale $s$ as $\sqrt{s}|H^*(s\omega_0)|$.

**3.** For a linear function $f(t)=t$,

$$W_f(s,\tau) = \frac{1}{\sqrt{s}} \int t h^*\left(\frac{t-\tau}{s}\right) dt$$
$$= s^{3/2} \int t h^*(t-\tau') dt = \frac{s^{3/2}}{j} \left.\frac{dH^*(\omega)}{d\omega}\right|_{\omega=0} \qquad (2.3.27)$$

Hence, if the wavelet $h(t)$ is regular and of order $n \geq 1$ so that its moments of order $n \geq 1$ is equal to zero and its derivatives of first-order is equal to zero at $\omega=0$, then the wavelet transform of $f(t)=t$ is equal to zero.

### 2.3.2.4.1 Wavelet Transform of Regular Signals

According to earlier discussion, the wavelet transform of a constant is zero. The wavelet transform of a linear signal is zero, when the wavelet has the first-order vanishing moment: $M_1=0$. The wavelet transform of a quadratic signal could be zero, when the wavelet has the first-and second-order vanishing moments: $M_1 = M_2 = 0$. The wavelet transform of a polynomial signal of degree *m* could be equal to zero, when the wavelet has the vanishing moments up to the order *n≥m*.

Thus, the wavelet transform is efficient for detecting singularities in signal and analyzing nonstationary, transient signal. For most functions the wavelet transforms have no closed analytical solutions and can be calculated only by numerical computer or by optical analog computer. The optical continuous wavelet transform is based on the explicit definition of the wavelet transform Equation (2.3.6) and implemented using a bank of optical wavelet transform filters as described in Section 2.3.1.1.1 in the Fourier plane in an optical correlator [14], [15], [16].

### 2.3.3 Discrete Wavelet Transform

The continuous wavelet transform is a mapping of a 1-D time signal into a 2-D timescale joint representation. The time bandwidth product of the continuous wavelet transform output is the square of that of the signal. For most applications, however, the goal of signal processing is to represent the signal efficiently with fewer parameters. The use of the discrete wavelet transform can reduce the time-bandwidth product of the wavelet transform output.

The discrete wavelet transform is understood as the continuous wavelets with the discrete scale and translation factors. The wavelet transform is then evaluated at discrete scales and translations. The discrete scale is expressed as $s = s_0^i$, where *i* is integer and $s_0 > 1$ is a fixed dilation step. The discrete translation factor is expressed as $\tau = k\tau_0 s_0^i$, where *k* is integer. The translation depends on the dilation $s_0^i$. The corresponding discrete wavelets are written as

$$\begin{aligned} h_{i,k}(t) &= s_0^{-i/2} h(s_0^{-i}(t - k\tau_0 s_0^i)) \\ &= s_0^{-i/2} h(s_0^{-i} t - k\tau_0) \end{aligned} \quad (2.3.28)$$

The discrete wavelet transform with the dyadic scaling factor with $s_0 = 2$ is effective in the computer implementation.

***Wavelet Frame***: With the discrete wavelet basis, a continuous function *f(t)* is transformed to a sequence of wavelet coefficients

$$W_f = \int f(t) h_{i,k}^*(t) dt = <f, h_{i,k}> \quad (2.3.29)$$

A raising questing for the discrete wavelet transform is how well the function $f(t)$ can be reconstructed from the discrete sequence of wavelet coefficients:

$$f(t) = A \sum_i \sum_k W_f(i,k) h_{i,k}(t) \qquad (2.3.30)$$

where $A$ is a constant that does not depend on $f(t)$. Obviously, if $s_0$ is close enough to 1 and $\tau_0$ is small enough, the set of wavelets approaches as continuous. The reconstruction (Equation 2.3.30) is then close to the inverse continuous wavelet transform. The signal reconstruction takes place without restrictive conditions other than the admissible condition on the wavelet $h(t)$. On the other hand, if the sampling is sparse, $s_0 = 2$ and $\tau_0 = 1$, the reconstruction (Equation 2.3.30) can be achieved only for some special choices of the wavelet $h(t)$.

Daubechies [17] has proven that the necessary and sufficient condition for the stable reconstruction of a function $f(t)$ from its wavelet coefficients $W_f(i, k)$ is that the energy, which is the sum of square moduli of $W_f(i, k)$, must lie between two positive bounds:

$$A \|f\|^2 \leq \sum_{j,k} |<f, h_{i,k}>|^2 \leq B \|f\|^2 \qquad (2.3.31)$$

where $\|f\|^2$ is the energy of $f(t)$, $A>0$, $B<\infty$ and $A, B$ are independent of $f(t)$. When $A=B$, the energy of the wavelet transform is proportional to the energy of the signal. This is similar to the energy conservation relation (Equation 2.3.16) of the continuous wavelet transform. When $A \neq B$, there is still some proportional relation between the energies of the signal and its wavelet transform. When Equation (2.3.31) is satisfied, the family of the wavelet basis functions $\{h_{i,k}(t)\}$ with $i, k \in Z$ is referred to as a frame and $A, B$ are termed frame bounds. The closer are $A$ and $B$, the more accurate is the reconstruction.

When $A=B$, the frame is tight and the discrete wavelets behave exactly like an orthonormal basis. When $A=B=1$ Equation (2.3.31) is simply the energy conservation equivalent to the Parseval relation of the Fourier transform. It is important to note that the same reconstruction works even when the wavelets are not orthogonal to each other.

When $A \neq B$ the reconstruction can still work exactly for the discrete wavelet transform if reconstruction uses the synthesis function basis, which is different from the decomposition function basis for analysis. The former constitute the dual frame of the later.

## 2.3.4 Applications of the Wavelet Transform

This section presents some popular applications of the wavelet transform for multiresolution transient signal analysis and detection, image edge detection, and compression.

### 2.3.4.1 Two-Dimensional Wavelet Transform

The wavelet transform can be easily extended to 2-D case for image processing applications. The wavelet transform of a 2-D image $f(x, y)$ is

$$W_f(s_x, s_y; u, v) = \frac{1}{\sqrt{s_x s_y}} \iint f(x, y) \psi\left(\frac{x-u}{s_x}; \frac{y-v}{s_y}\right) dx dy \qquad (2.3.32)$$

that is a four-dimensional function. It is reduced to a set of two-dimension functions of $(u, v)$ with different scales, when the scale factors $s_x = s_y = s$. When $\psi(x, y) = \psi(r)$ with $r = (x^2+y^2)^{1/2}$, the wavelets are isotropic and have no selectivity for spatial orientation. Otherwise, the wavelet can have particular orientation. The wavelet can also be a combination of the 2-D wavelets with different particular orientations, so that the 2-D wavelet transform has orientation selectivity.

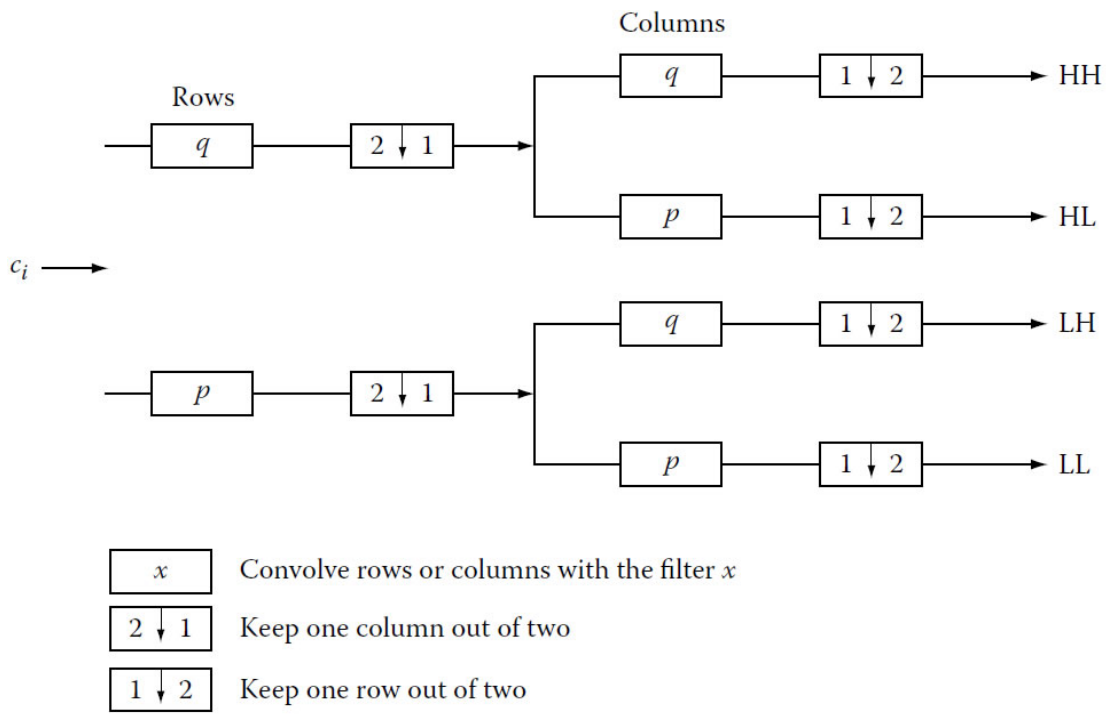

Fig 2.1. Schematic two-dimensional wavelet decomposition with quadrature mirror low-pass and high-pass filters $p(n)$ and $q(n)$ [11].

At each resolution the pair of the 1-D low-pass and high-pass filters are first applied to each row of the image that result in a horizontal approximation image and a horizontal detail image. Then the pair of the 1-D filters are applied to each column of the two horizontally filtered images. The down-sampling by two is downsampling result in four subband images: (LL) for the low-pass filtered both horizontally and vertically image, (HH) for the highpass filtered both horizontally and vertically image, (LH) for lowpass filtered in horizontal direction and high-pass filtered in vertical direction image and (HL) for high-pass filtered in vertical direction and high-pass filtered in horizontal direction image, as shown in Figure 2.1 [18].

All the four images have the half size of the input image. The detail images (LH), (HL) and (HH) are now put in three respective quadrants as shown in Figure 2.2. The image (LL) is the approximation image in both horizontal and vertical directions and is down-sampled in both directions. Then, whole process of two-step filtering and down-sampling is applied again to the image (LL) in this lower resolution level. The iteration can continue many times until, for instance, the image (LL) has only a size of 2X2. Figure 2.2 shows a disposition of the detail images (LH), (HL) and (HH) at three resolution levels (1, 2, 3) and the approximation image (LL) at the fourth low resolution level (4).

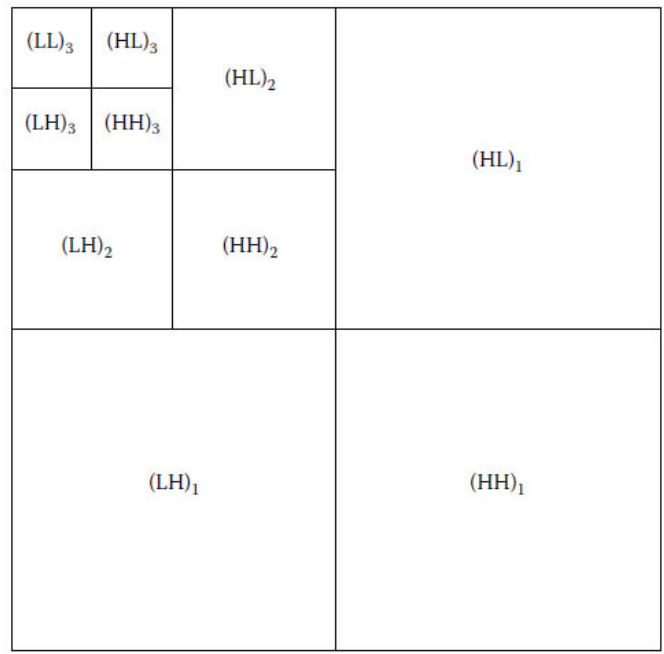

Fig 2.2: Presentation of the two-dimensional wavelet decomposition and high-pass filters $p(n)$ and $q(n)$ [11].

If the original image has $L^2$ pixels at the resolution $i$=0, then each image (LH), (HL) and (HH) at resolution level $i$ has $(L/2^i)^2$ pixels ($i$>0) The total number of pixels of the orthonormal wavelet representation is therefore still equal to $L^2$, as shown in Figure 2.2.

The dyadic wavelet transform does not increase the volume of data. This is owing to the orthonormality of the discrete wavelet decomposition.

*2.3.4.2 Image Compression*

Image compression is to use fewer bits to represent the image information for different purposes, such as image storage, image transmission, and feature extraction. The general idea behind is to remove the redundancy in an image.

A popular method for image compression for removing the spatial redundancy is transform coding that represents the image in the transformation basis such that the transformation coefficients are decorrelated. The multiresolution wavelet decomposition is projections onto subspaces spanned by scaling function basis and the wavelet basis. The projections on the scaling function basis yield approximations of the signal and the projections on the wavelet basis yield the differences between the approximations at two adjacent resolution levels. Therefore, the wavelet detail images are decorrelated and can be used for image compression. Indeed, the detail images obtained from the wavelet transform consist of edges in the image. There is only little correlation among the values on pixels in the edge images.

One example of image compression applications is the grayscale fingerprint image compression using wavelet transform [19]. The fingerprint images are captured as 500 pixels per inch and 256 gray levels. The wavelet sub band decomposition is accomplished by the tree algorithm described by Figure 2.1. The dominant ridge frequency in fingerprint images is in roughly $\omega=\pi/8$ up to $\omega=\pi/4$ bands. Because the wavelet decomposition removes the correlation among image pixels, only the wavelet coefficients with large magnitude are retained.

Most wavelet transform coefficients are equal or close to zero in the regions of smooth image intensity variation. After a thresholding on the wavelet coefficients the retained coefficients are subsequently coded according to a scalar quantizer and are mapped to a set of 254 symbols for Huffman encoding using the classical image coding technique.

The thresholding and the Huffman coding can achieve high compression ratio. The decoder can now reconstruct approximations of the original images by performing inverse wavelet transform using the low-pass and high-pass filter analysis. After compression at 20:1, the reconstructed images conserve the ridge features: ridge ending or bifurcations that are definitive information useful for determination.

## 2.4 Chaos Function.

Chaos functions have been used mainly to develop the mathematical models of non-linear systems. There are several interesting properties of the chaos functions [20]. These functions generate iterative values which are completely random in nature but limited between bounds.

Convergence of the iterative values after any value of iterations can never be seen. Chaos functions have extreme sensitivity to the initial conditions. Three chaos functions are discussed below

(a) The first chaos function is the logistic map [20], [21], [22] and is defined as:

$$f(x) = p.x.(1-x) \quad (2.4.1)$$

This function is bounded for $0 < p < 4$ and can be written in the iterative form as:

$$x_{n+1} = p.x_n.(1-x_n) \quad (2.4.2)$$

with '$x_0$' as the initial value. This is also known as the seed value for the chaos function.

(b) The second chaos function is the tent map [20],[21], [22] and is defined as:

$$f(x) = \begin{cases} a.x & 0 \le x_0 \le 0.5 \\ a.(1-x) & 0.5 \le x_0 \le 1 \end{cases} \quad (2.4.3)$$

This function is bounded for $0 < a \le 2$ and can be written in the iterative form as:

$$x_{n+1} = \begin{cases} a.x_n & 0 \le x_0 \le 0.5 \\ a.(1-x_n) & 0.5 \le x_0 \le 1 \end{cases} \quad (2.4.4)$$

with '$x_0$' as the initial value.

(c) The third chaos function is the Kaplan–Yorke map [20], [21], [22] and is defined as:

$$\left. \begin{array}{l} f(x) = a.x.\mathrm{mod}\,1 \quad 0 \le x_0 \le 0.5 \\ f(y) = by + \cos(4\pi x) \quad 0.5 \le x_0 \le 1 \end{array} \right\} \quad (2.4.5)$$

This is bounded for $0 \le a \le 2$ and $0 \le b \le 1$ and can be written in the iterative form as:

$$x_{n+1} = a.x_n \,\mathrm{mod}\,1 \quad (2.4.6)$$

$$y_{n+1} = by_n + \cos(4\pi x_n) \qquad (2.4.7)$$

with '$x_0$' as the initial value. These chaos functions are used to generate random phase masks. Logistic map and tent map are one-dimensional chaos functions and the Kaplan–Yorke map is a two-dimensional chaos function. For the two-dimensional chaos function, two seed values are required to generate the CRPM (Chaotic random phase mask).

## 2.5 Random phase Mask.

The process of making a random phase mask is described as follows. First, a chaotic map $S(x, y)$ is generated by introducing some input parameters to the map. These input parameters are called seed values and the size of the map respectively. A typical chaotic map of size $M \times N$ pixels generated based on the mentioned technique is shown in fig 2.3 and its chaotic behavior is plotted in fig 2.4.

The chaotic map obtained in fig 2.3 is real. The random chaotic map is generated from the chaotic map using:

$$\phi(x, y) = \exp\left(i\frac{\pi}{2}S(x, y)\right) \qquad (2.4.8)$$

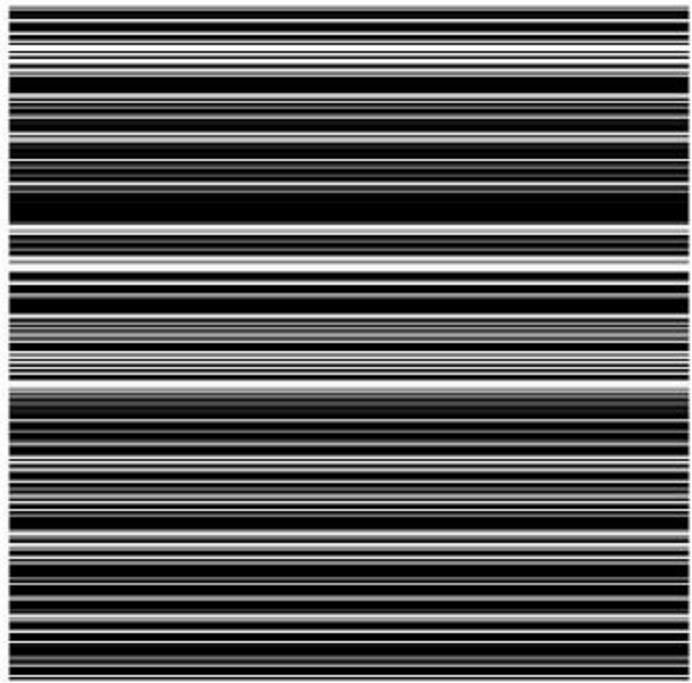

Fig 2.3. A chaotic map of size 256X256.

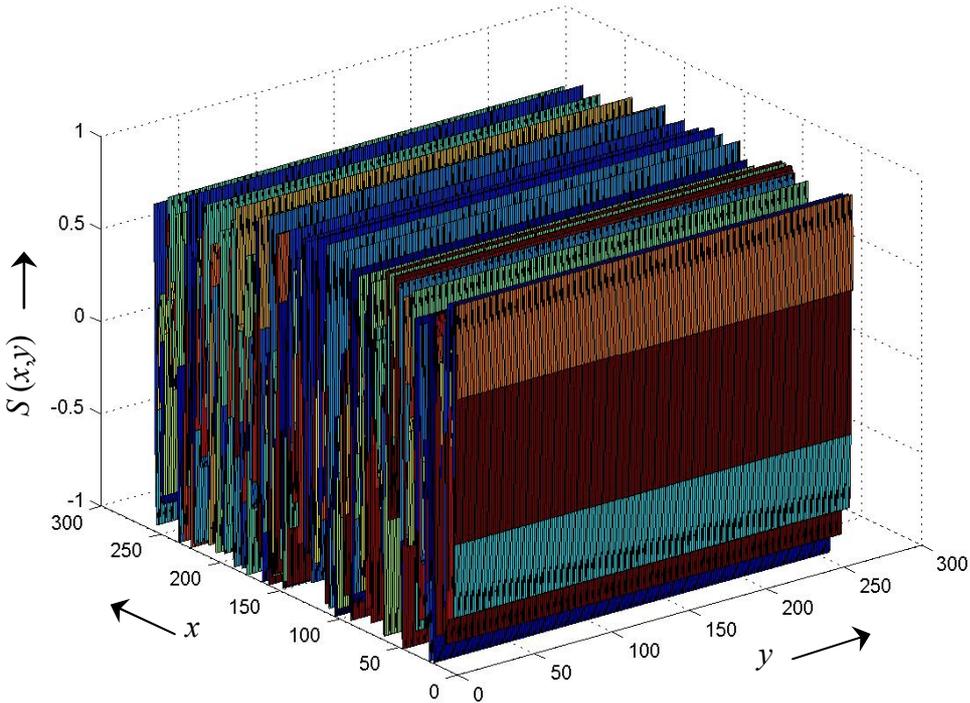

Fig 2.4. Behaviour of the chaotic map shown in Fig 2.3.

<div style="text-align: right;">Chapter – 3</div>

# ANALYTICAL STUDY OF EXISTING METHODS

This chapter discusses three different methods, used for color image encryption and decryption process proposed in [1] and [2]. For further discussion a 2-*D* color image will be denoted by a 2-*D* function *f*(*x*, *y*) where *x*, *y* are the spatial coordinates of the image. Therefore an image I of size *M* X *N* can be represented as:

$$I = \sum_{x=1}^{M} \sum_{y=1}^{N} f(x,y) \qquad (3.1)$$

where *f*(*x*, *y*) is further represented as follows:

$$f(x,y) = \sum_{i=1}^{3} f_i(x,y) \qquad (3.2)$$

Here, *i* denote the index of primary color component *i.e.* $f_1(x,y)$, $f_2(x,y)$ and $f_3(x,y)$ which correspond to three primary color channels *viz*. Red, Green and Blue respectively. Without the loss of generality *M=N* can be chosen for analysis of algorithms.

**3.1 Algorithm 3.1: Image encryption using *FRT*:** As proposed in [1], the original image $f(x,y)$ is first segregated into its three primary color channels (R, G and B). Each of these channels is then multiplied by a random phase function $\phi_1(x,y)$ and is subsequently Fourier transformed using a 2-*D* Fourier transform using parameters α and β. In the next step, the Fourier transformed data is multiplied by another random phase function $\phi_2(x,y)$, which is statistically independent of the first random phase function. Another 2-*D* Fourier transform with parameters γ and δ is now performed on each of the primary color channels obtained after previous step, thus, producing three encrypted primary color channels. These channels, when merged together, produce an encrypted image $I_1 = g(x, y)$.

Decryption process is the inverse of the encryption process. Encrypted image is first segregated into three primary color channels *i.e.* R, G and B denoted by $g_1(x,y)$, $g_2(x,y)$ and $g_3(x,y)$. These channels are inverse Fourier transformed using a 2-*D* Fourier transforms of suitable fractional order. Each of these channels is then multiplied by $conj(\phi_2(x,y))$ and inverse Fourier transformed again, using a 2-*D* Fourier transform of suitable fractional order.

A multiplication by $conj(\phi_1(x,y))$ now produces R, G and B channels of a decrypted image, say $I' = f'(x,y)$.

**Computation complexity:**

**Encryption:** Let $I$ be an image of size $N$ X $N$. Encryption process starts with three primary color channels of $I$ and involves a sequence of following steps:

**(i) Multiplication by $\phi_1(x,y)$:** This multiplication takes place element by element between R, G and B components of original image and 2-D matrix of random phase function $\phi_1(x,y)$, which involves $N^2$ multiplication operations. Therefore its computation complexity is $O(N^2)$.

**(ii) 2-D Fourier transform of the output of step (i):** Computation of 2-D Fourier transform is divided into two phases:

  **a)** Generation of Fourier transform matrix.
  **b)** Computation of 2-D Fourier transform of input matrix.

Computation complexity of this process is evaluated to $O(2N^3 + N\log_2 N)$ [23].

**(iii) Multiplication by $\phi_2(x,y)$:** Similar to the step (i), the computation complexity of this step is also $O(N^2)$.

**(iv) 2-D Fourier transform of the output of step (iii):** Similar to the step (ii) Computation complexity of this process is evaluated to $O(2N^3 + N\log_2 N)$.

Therefore, time complexity of encryption process of Algorithm 3.1 (step (i) – (iv)) can be represented as follows:

$$T_{encryption} = O(N^2) + O(2N^3 + N\log_2 N) + O(N^2) + O(2N^3 + N\log_2 N)$$
$$= O(4N^3 + 2N^2 + 2N\log_2 N)$$
$$\Rightarrow T_{encryption} \approx O(4N^3)$$

It is to be noted here that lower order terms can be neglected while evaluating asymptotic upper bound on time complexity [24] [25].

**Decryption:** Let $I_1 = g(x,y)$ be the outcome of encryption process and the size of $I_1$ be $N$X$N$. Decryption process starts with three primary color channels of $g$ and involves a sequence of following steps:

**(i) 2-*D* inverse Fourier transform of R,G & B of $I_1$** : Computation of 2-*D* inverse Fourier transform is divided into two phases :

    **a)** Generation of inverse Fourier transform matrix.

    **b)** Computation of 2-*D* Fourier transform of input matrix.

Computation complexity of this process is evaluated to $O(2N^3 + N\log_2 N)$ [23].

**(ii) Multiplication by $conj(\phi_2(x,y))$**: This multiplication takes place element by element between R, G and B components of output of step (i) and 2-*D* matrix of random phase function $conj(\phi_2(x,y))$, which involves $N^2$ multiplication operations. Therefore its computation complexity is $O(N^2)$.

**(iii) 2-*D* Fourier transform of the output of step (ii)**: Similar to the step (i) Computation complexity of this process is evaluated to $O(2N^3 + N\log_2 N)$.

**(iv) Multiplication by $conj(\phi_1(x,y))$** : Similar to the step (i), the computation complexity of this step is also $O(N^2)$.

Therefore, time complexity of decryption process of Algorithm 3.1 (step (i) – (iv)) can be represented as follows:

$$T_{decryption} = O(N^2) + O(2N^3 + N\log_2 N) + O(N^2) + O(2N^3 + N\log_2 N)$$
$$= O(4N^3 + 2N^2 + 2N\log_2 N)$$
$$\Rightarrow T_{decryption} \approx O(4N^3)$$

Thus, the computation complexity of algorithm 3.1 is evaluated to:

$$T_{3.1} = T_{encryption} + T_{decryption} = O(8N^3)$$

**3.2 Algorithm 3.2 & 3.3: Image encryption using *FRT* and chaos:**

Algorithm 3.2 and 3.3 follow a sequence of operations, similar to algorithm 3.1 but with a little difference. As discussed in section 2.3, chaos functions generate iterative values which are completely random in nature but limited between bounds [2]. A color image encryption and decryption method was proposed in [2]. For further discussion only **(i)** Logistic map and **(ii)** Kaplan-Yorke map are chosen due to the reason that the Logistic map and Tent map are one dimensional chaos functions whereas the Kaplan-Yorke map is a two-dimensional chaos function. These chaos functions are used to generate the CRPM (chaotic random phase mask). In case of one-dimensional chaos functions (Logistic map and Tent map) only one seed value is needed whereas in case of a two-dimensional chaos functions (Kaplan-Yorke map) two seed values are needed to generate the CRPM.

The original image *I* denoted by $f(x,y)$ is composed of three primary color channels *viz*. R, G and B. Each of these channels is multiplied by the first CRPM and is subsequently Fourier transformed using a 2-*D* Fourier transform. In the next step, the Fourier transformed coefficients are multiplied by another CRPM, which is generated by a different seed value than the first CRPM. Another 2-*D* Fourier transform is now performed on each of the primary color channels obtained after previous step, thus, producing three encrypted primary color channels. These channels, when merged together, produce an encrypted image $I_1 = g(x, y)$.

For further discussion, Algorithm 3.2 refers to encryption and decryption method using fractional Fourier transform and Logistic map and Algorithm 3.3 refers to encryption and decryption method using fractional Fourier transform and Kaplan-Yorke map. In both the methods, chaotic functions are used to generate the first and second CRPM denoted by $\phi_1(x,y)$ and $\phi_2(x,y)$ respectively. Following sections discuss the computation complexity of Algorithm 3.2 and Algorithm 3.3 separately.

### 3.2.1 Computation complexity of *Algorithm* 3.2:

**Encryption:** Let *I* be an image of size *N* X *N*. Encryption process starts with three primary color channels of *I* and involves a sequence of following steps:

(i) **Multiplication by first CRPM** $\phi_1(x,y)$ : This multiplication takes place in two phases (a) Generation of $\phi_1(x,y)$ using logistic map, which is a linear process having $O(N^2)$ computation complexity and (b) element by element multiplication of $\phi_1(x,y)$ with $f_i(x,y)$. The computation complexity of this phase is $O(N^2)$. Therefore computation complexity of step (i) is $O(2N^2)$.

(ii) **2-*D* Fourier transform of the output of step (i)**: Similar to section 3.1, computation complexity of this step $O(2N^3 + N\log_2 N)$.

(iii) **Multiplication by second CRPM** $\phi_2(x,y)$: Similar to the step (i), the computation complexity of this step is also $O(2N^2)$.

**(iv) 2-D Fourier transform of the output of step (iii)**: Similar to the step (ii) Computation complexity of this process is evaluated to $O(2N^3 + N\log_2 N)$.

Therefore, computation complexity of encryption process of Algorithm 3.2 (step (i) – (iv)) can be represented as follows:

$$T_{encryption} = O(2N^2) + O(2N^3 + N\log_2 N) + O(2N^2) + O(2N^3 + N\log_2 N)$$
$$= O(4N^3 + 4N^2 + 2N\log_2 N)$$
$$\Rightarrow T_{encryption} \approx O(4N^3)$$

**Decryption:** Let $I_1 = g(x, y)$ be the outcome of encryption process. Size of $I_1$ is $N \times N$. Decryption process starts with three primary color channels of $g$ and involves a sequence of following steps:

**(i) 2-D inverse Fourier transform of R,G & B of $I_1$** : Similar to section 3.1 computation complexity of this step is $O(2N^3 + N\log_2 N)$.

**(ii) Multiplication by $conj(\phi_2(x,y))$**: This multiplication takes place element by element between R, G and B components of output of step (i) and 2-D matrix of random phase function $conj(\phi_2(x,y))$, which involves $N^2$ multiplication operations. Therefore its computation complexity is $O(2N^2)$, out of which $O(N^2)$ time is required to produce $conj(\phi_2(x,y))$ from $\phi_2(x,y)$.

**(iii) 2-D Fourier transform of the output of step (ii)**: Similar to the step (i) Computation complexity of this step is evaluated to $O(2N^3 + N\log_2 N)$.

**(iv) Multiplication by $conj(\phi_1(x,y))$** : Similar to the step (ii), the computation complexity of this step is also $O(2N^2)$.

Therefore, computation complexity of decryption process of algorithm 3.2 (step (i) – (iv)) can be represented as follows:

$$T_{decryption} = O(2N^2) + O(2N^3 + N\log_2 N) + O(2N^2)O(2N^3 + N\log_2 N)$$
$$= O(4N^3 + 4N^2 + 2N\log_2 N)$$
$$\Rightarrow T_{decryption} \approx O(4N^3)$$

Thus, the computation complexity of Algorithm 3.2 is evaluated to:

$$T_{3.2} = T_{encryption} + T_{decryption} = O(8N^3)$$

**3.2.2 Computation complexity of *Algorithm* 3.3:**

As discussed in the beginning of section 3.2, computation of algorithm 3.3 is similar to algorithm 3.2 except the difference that algorithm 3.3 uses Kaplan-Yorke function to generate $\phi_1(x, y)$ and $\phi_2(x, y)$. Therefore, computations using Kaplan-Yorke function only may cause difference in computation times of algorithm 3.2 and algorithm 3.3

**Encryption:** Let *I* be an image of size *N* X *N*. Encryption process starts with three primary color channels of *I* and involves a sequence of following steps:

**(i) Multiplication by first CRPM $\phi_1(x, y)$** : This multiplication takes place in two phases **(a)** Generation of $\phi_1(x, y)$ using Kaplan-Yorke map, which is a linear process having $O(N^2)$ computation complexity and **(b)** element by element multiplication of $\phi_1(x, y)$ with $f_i(x, y)$ which is also $O(N^2)$. Therefore computation complexity of step (i) is $O(2N^2)$.

**(ii) 2-*D* Fourier transform of the output of step (i)**: Similar to section 3.2.1, computation complexity of this step $O(2N^3 + N\log_2 N)$.

**(iii) Multiplication by second CRPM $\phi_2(x, y)$** : Similar to the step (i), the computation complexity of this step is also $O(2N^2)$.

**(iv) 2-*D* Fourier transform of the output of step (iii)**: Similar to the step (ii) computation complexity of this process is $O(2N^3 + N\log_2 N)$.

Therefore, computation complexity of encryption process of algorithm 3.3 (step (i) – (iv)) can be represented as follows:

$$T_{encryption} = O(2N^2) + O(2N^3 + N\log_2 N) + O(2N^2) + O(2N^3 + N\log_2 N)$$
$$= O(4N^3 + 4N^2 + 2N\log_2 N + 2N)$$
$$\Rightarrow T_{encryption} \approx O(4N^3)$$

**Decryption:** Let $I_1 = g(x, y)$ be the outcome of encryption process. Size of $I_1$ is *N*X*N*. Decryption process starts with three primary color channels of *g* and involves a sequence of following steps:

**(i) 2-*D* inverse Fourier transform of R, G & B of $I_1$**: Similar to section 3.2.1 computation complexity of this step is $O(2N^3 + N\log_2 N)$.

**(ii) Multiplication by $conj(\phi_2(x, y))$**: Similar to section 3.2.1, computation complexity of this step is $O(2N^2)$.

**(iii) 2-D Fourier transform of the output of step (ii)**: Similar to the step (i), computation complexity of this step is evaluated to $O(2N^3 + N\log_2 N)$.

**(iv) Multiplication by** $conj(\phi_1(x,y))$ : Similar to the step (ii), the computation complexity of this step is also $O(2N^2)$.

Therefore, computation complexity of decryption process of method 1 (step (i) – (iv)) can be represented as follows:

$$T_{decryption} = O(2N^2) + O(2N^3 + N\log_2 N) + O(2N^2) + O(2N^3 + N\log_2 N)$$

$$= O(4N^3 + 4N^2 + 2N\log_2 N + 2N)$$

$$\Rightarrow T_{decryption} \approx O(4N^3)$$

Thus, the computation complexity of algorithm 3.3 is evaluated to:

$$T_{3.3} = T_{encryption} + T_{decryption} = O(8N^3).$$

## PROPOSED ALGORITHMS.

Chapter 3 discussed that a color image $I = f(x,y)$ of size $M$ X $N$ consists of three primary color channels *viz*. RED, GREEN and BLUE. For simplicity the image size is assumed to be $N$ X $N$. This chapter proposes three image-encryption and decryption algorithms based on the Discrete Wavelet Transform, fractional Fourier transform and chaotic logistic-map and Kaplan-Yorke map. Each of these methods follows a similar sequence of operations but differing in their implementation. An overview of the encryption and decryption process is as follows.

As shown in fig 4.1, initially, the RED, GREEN and BLUE channels of the original image are segregated. Rests of the following operations are applied concurrently to these channels. Initially, the $DWT_2$ operation is performed over each channel to generate:

$$DWT_2\{f_i(x,y)\}, \quad 1 \leq i \leq 3$$

This distribution is encoded by the first CRPM which is mathematically expressed as the phase function $\exp\left(j\frac{\pi}{2}S_1(x,y)\right)$, where $S_1(x,y)$ is the random number sequence generated by the chaos function. The first 2-D *FRT* operation is then performed over this to generate:

$$F_{\gamma,\delta}\left\{DWT_2\{f_i(x,y)\} * \exp\left(j\frac{\pi}{2}S_1(x,y)\right)\right\}$$

where γ, δ are the fractional orders of the first 2-D *FRT* and * denote the element by element multiplication between two matrices of same order. The resultant matrix is encoded by the second CRPM which is mathematically expressed as the phase function $\exp\left(j\frac{\pi}{2}S_2(x,y)\right)$, where $S_2(x,y)$ is the random number sequence generated by the chaos function for a different seed value than the first CRPM. . The second *FRT* operation is then performed over this to give us:

$$F_{\alpha,\beta}\left\{F_{\gamma,\delta}\left\{DWT_2\{f_i(x,y)\}\exp\left(j\frac{\pi}{2}S_1(x,y)\right)\right\} * \exp\left(j\frac{\pi}{2}S_2(x,y)\right)\right\}$$

where α, β are the fractional orders of the second 2-D *FRT*. Each channel is now operated with $IDWT_2$ to produce R, G and B channel of the encrypted image. These channels are merged to produce the encrypted image $g(x, y)$ as per the following formulae:

$$I_1 = g(x,y) = IDWT_2\left\{F_{\alpha,\beta}\left\{F_{\gamma,\delta}\left\{DWT_2\{f_i(x,y)\} * \exp\left(j\frac{\pi}{2}S_1(x,y)\right)\right\} * \exp\left(j\frac{\pi}{2}S_2(x,y)\right)\right\}\right\}$$

………… (4.1)

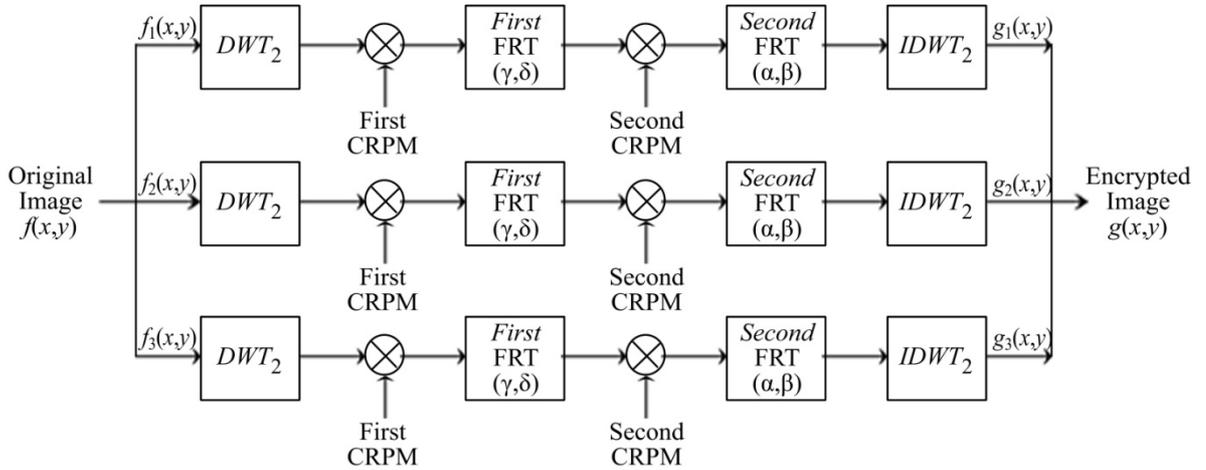

Fig 4.1. Encryption process using $DWT_2$ in proposed algorithm.

The decryption process as shown in fig 4.2 is the inverse of the encryption process. The $DWT_2$ operation is performed over $g(x,y)$, the encrypted image, to generate:

$$DWT_2\{g(x,y)\}$$

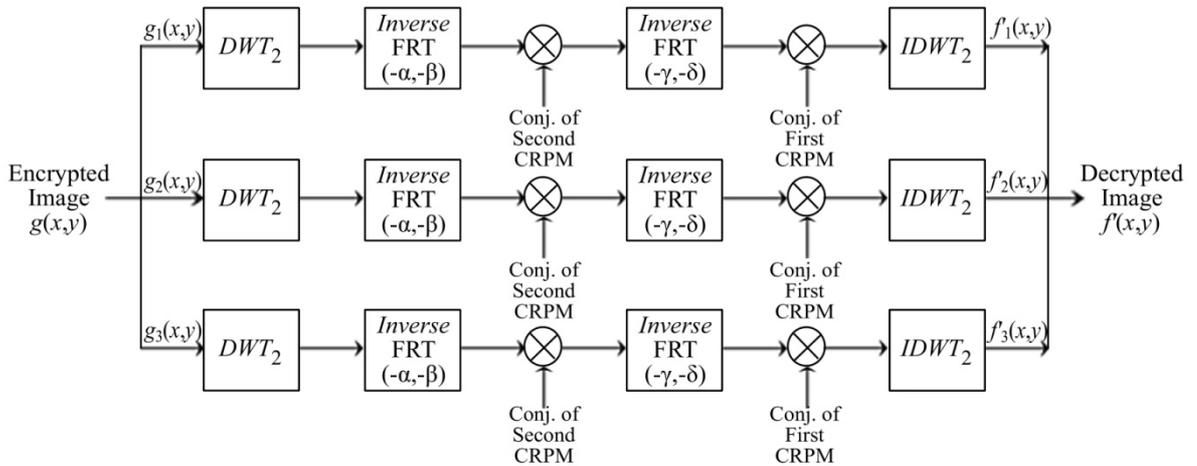

Fig 4.2. Decryption process using $IDWT_2$ in proposed algorithm.

The first inverse $FRT$ (of order -α, -β) is now applied and followed by a multiplication by conjugate of second CRPM, thus generating:

$$F_{-\alpha,-\beta}\{DWT_2\{g(x,y)\}\} * conj\left(\exp\left(j\frac{\pi}{2}S_2(x,y)\right)\right)$$

On the output obtained, the second *FRT* (of order -γ, -δ) is performed and then multiplication by the conjugate of first CRPM takes place. The decrypted image from this outcome is now obtained by performing $IDWT_2$ as follows:

$$f'(x,y) = IDWT_2\left\{F_{-\gamma,-\delta}\left\{F_{-\alpha,-\beta}\{DWT_2\{g(x,y)\}\} * conj\left(\exp\left(j\frac{\pi}{2}S_2(x,y)\right)\right)\right\} * conj\left(\exp\left(j\frac{\pi}{2}S_1(x,y)\right)\right)\right\}$$

… … … … (4.2)

**4.1 Algorithm 4.1:** *Image encryption and decryption using $DWT_2$ and FRT.*

This algorithm uses $DWT_2$ and $F_{\alpha,\beta}$ in encryption process and $IDWT_2$ and $F_{-\alpha,-\beta}$ in decryption process as per the general encryption and decryption schemes shown in fig. 4.1 and fig. 4.2. Both of the random phase functions are generated as a 2-D sequence of random numbers and they are not chaos based in this algorithm.

**Computation complexity:**

Let the input image *I* be of size *N* x *N*. Analysis of the algorithm is divided into two phases *viz*. Encryption and Decryption.

**Encryption:** Image encryption process involves following steps:

1. Application of $DWT_2$ on the primary color channels $f_i(x, y)$ of original image.
2. Encoding by first random phase function.
3. Application of first 2-D *FRT*.
4. Encoding by second random phase function.
5. Application of second 2-D *FRT*.
6. Application of $IDWT_2$ on the R, G, and B channels obtained after step 6.

In step 1, each of the primary color channels of original image is operated using $DWT_2$, which is an $O(N)$ process. Thus step 1 takes $O(N)$ time and produces an output image of size $\frac{N}{2} \times \frac{N}{2}$. Therefore, steps 2 to 6 are to be applied on a smaller sized image than the original.

Computation of 2-D *FRT* of an image of size *N* x *N* is a $O(2N^3 + N\log_2 N)$ [23] process. Also, the generation and multiplication of first and second random phase function is an $O(2N^2)$ process. But now, as the input image size for steps 2 to 6 has reduced to $\frac{N}{2} \times \frac{N}{2}$,

**(i)** Step 2 takes $O\left(2\left(\frac{N}{2}\right)^2\right) = O\left(\frac{N^2}{2}\right)$ computation time.

**(ii)** Step 3 takes $O\left(2\left(\frac{N}{2}\right)^3 + \frac{N}{2}\log_2\left(\frac{N}{2}\right)\right) = O\left(\frac{N^3}{4} + \frac{N}{2}\log_2\left(\frac{N}{2}\right)\right)$ time.

Similar to step 2 and 3, step 4 and 5 also take $O\left(\frac{N^2}{2}\right)$ and $O\left(\frac{N^3}{4} + \frac{N}{2}\log_2\left(\frac{N}{2}\right)\right)$ computation time, respectively. Step 6 involves computation of inverse wavelet transform, which is also a $O(N)$ function. Therefore, the computation complexity of encryption is:

$$T_{encryption} = O(N) + O\left(\frac{N^2}{2}\right) + O\left(\frac{N^3}{4} + \frac{N}{2}\log_2\left(\frac{N}{2}\right)\right) + O\left(\frac{N^2}{2}\right)$$

$$+ O\left(\frac{N^3}{4} + \frac{N}{2}\log_2\left(\frac{N}{2}\right)\right) + O(N)$$

$$= O\left(\frac{N^3}{2} + N^2 + N\log_2\left(\frac{N}{2}\right) + 2N\right)$$

$$\Rightarrow T_{encryption} = O\left(\frac{N^3}{2}\right)$$

**Decryption:** Image decryption process involves following steps:

1. Application of $DWT_2$ on the primary color components $g_i(x,y)$ of encrypted image.
2. Application of 2-D inverse *FRT*.
3. Decoding by conjugate of second CRPM.
4. Application of 2-D inverse *FRT*.
5. Decoding by conjugate of first CRPM.
6. Application of $IDWT_2$ on the R, G, and B channels obtained after step 5.

In step 1, each of the primary color channels of encrypted image is operated using $DWT_2$, which is an $O(N)$ process. Thus step 1 takes $O(N)$ time and produces an output encrypted image of size $\frac{N}{2} \times \frac{N}{2}$. Therefore, steps 2 to 6 are to be applied on a smaller sized image than the original encrypted image.

**(i)** Step 2 takes $O\left(2\left(\frac{N}{2}\right)^3 + \frac{N}{2}\log_2\left(\frac{N}{2}\right)\right) = O\left(\frac{N^3}{4} + \frac{N}{2}\log_2\left(\frac{N}{2}\right)\right)$ time.

**(ii)** Step 3 takes $O\left(2\left(\frac{N}{2}\right)^2\right) = O\left(\frac{N^2}{2}\right)$ time.

Similar to step 2 and 3, step 4 and 5 also take $O\left(\frac{N^3}{4}+\frac{N}{2}\log_2\left(\frac{N}{2}\right)\right)$ and $O\left(\frac{N^2}{2}\right)$ computation time, respectively. Step 6 involves computation of inverse wavelet transform, which is an $O(N)$ process. Therefore, the computation complexity of decryption is:

$$T_{decryption} = O(N) + O\left(\frac{N^2}{2}\right) + O\left(\frac{N^3}{4}+\frac{N}{2}\log_2\left(\frac{N}{2}\right)\right) + O\left(\frac{N^2}{2}\right)$$

$$+ O\left(\frac{N^3}{4}+\frac{N}{2}\log_2\left(\frac{N}{2}\right)\right) + O(N)$$

$$= O\left(\frac{N^3}{2} + N^2 + N\log_2\left(\frac{N}{2}\right) + 2N\right)$$

$$\Rightarrow T_{decryption} = O\left(\frac{N^3}{2}\right)$$

Thus, the computation complexity of algorithm 4.1 is evaluated to:

$$T_{4.1} = T_{encryption} + T_{decryption} = O(N^3)$$

### 4.2 Algorithm 4.2: *Image encryption and decryption using DWT₂, FRT and logistic map.*

This algorithm is similiar to the Algorithm 4.1 except the fact that both of the random phase functions are generated using Chaotic Logistic map (as discussed in section 2.5).

**Computation complexity:**

Similar to the Algorithm 4.1, estimation of computation time of Algorithm 4.2 is divided into following two phases:

**Encryption:** Image encryption process involves following steps:

1. Application of $DWT_2$ on the primary color channels $f_i(x,y)$ of original image.
2. Encoding by first random phase function.
3. Application of first 2-D FRT.
4. Encoding by second random phase function.
5. Application of second 2-D FRT.
6. Application of $IDWT_2$ on the R, G, and B channels obtained after step 5.

For an input image of size $NXN$, step 1 is an $O(N)$ process which produces $\frac{N}{2} \text{X} \frac{N}{2}$ image. Thus, input for step 2 to 6 is an $\frac{N}{2} \text{X} \frac{N}{2}$ image matrix. Similar to section 4.1, computation time of step 2 and step 4 is $O\left(\dfrac{N^2}{2}\right)$ and that of step 3 and 5 is $O\left(\dfrac{N^3}{4}+\dfrac{N}{2}\log_2\left(\dfrac{N}{2}\right)\right)$. Step 6 involves computation of inverse wavelet transform, which is an $O(N)$ function. Therefore, the computation complexity of encryption is:

$$T_{encryption} = O(N) + O\left(\dfrac{N^2}{2}\right) + O\left(\dfrac{N^3}{4}+\dfrac{N}{2}\log_2\left(\dfrac{N}{2}\right)\right) + O\left(\dfrac{N^2}{2}\right)$$

$$+ O\left(\dfrac{N^3}{4}+\dfrac{N}{2}\log_2\left(\dfrac{N}{2}\right)\right) + O(N)$$

$$= O\left(\dfrac{N^3}{2} + N^2 + N\log_2\left(\dfrac{N}{2}\right) + 2N\right)$$

$$\Rightarrow T_{encryption} = O\left(\dfrac{N^3}{2}\right)$$

**Decryption:** Image decryption process involves following steps:

1. Application of $DWT_2$ on the primary color components $g_i(x, y)$ of encrypted image.
2. Application of 2-D inverse *FRT*.
3. Decoding by conjugate of second CRPM.
4. Application of 2-D inverse *FRT*.
5. Decoding by conjugate of first CRPM.
6. Application of $IDWT_2$ on the R, G, and B channels obtained after step 5.

For an input image of size $N\mathrm{X}N$, step 1 is an $O(N)$ process which produces $\dfrac{N}{2}\mathrm{X}\dfrac{N}{2}$ image. Thus, input for step 2 to 6 is an $\dfrac{N}{2}\mathrm{X}\dfrac{N}{2}$ image matrix. Similar to section 4.1, computation time of step 2 and step 4 is $O\left(\dfrac{N^2}{2}\right)$ and that of step 3 and 5 is $O\left(\dfrac{N^3}{4}+\dfrac{N}{2}\log_2\left(\dfrac{N}{2}\right)\right)$. Step 6 involves computation of inverse wavelet transform, which is an $O(N)$ function. Therefore, the computation complexity of decryption is:

$$T_{deryption} = O(N) + O\left(\dfrac{N^2}{2}\right) + O\left(\dfrac{N^3}{4}+\dfrac{N}{2}\log_2\left(\dfrac{N}{2}\right)\right) + O\left(\dfrac{N^2}{2}\right)$$

$$+ O\left(\frac{N^3}{4} + \frac{N}{2}\log_2\left(\frac{N}{2}\right)\right) + O(N)$$

$$= O\left(\frac{N^3}{2} + N^2 + N\log_2\left(\frac{N}{2}\right) + 2N\right)$$

$$\Rightarrow T_{decryption} = O\left(\frac{N^3}{2}\right)$$

Thus, the computation complexity of algorithm 4.2 is evaluated to:

$$T_{4.2} = T_{encryption} + T_{decryption} = O(N^3)$$

## 4.3 Algorithm 4.3: *Image encryption and decryption using DWT$_2$, FRT and Kaplan-Yorke map.*

This algorithm is similiar to the Algorithm 4.1 except the fact that both of the random phase functions are generated using Kaplan-Yorke map, which is a two-dimensional function (as discussed in section 2.5).

**Computation complexity:**

Similar to the Algorithm 4.1, estimation of computation time of Algorithm 4.3 is divided into following two phases:

**Encryption:** Image encryption process involves following steps:

1. Application of *DWT$_2$* on the primary color channels $f_i(x,y)$ of original image.
2. Encoding by first random phase function.
3. Application of first 2-D *FRT*.
4. Encoding by second random phase function.
5. Application of second 2-D *FRT*.
6. Application of *IDWT$_2$* on the R, G, and B channels obtained after step 5.

For an input image of size *NXN*, step 1 is an $O(N)$ process which produces $\frac{N}{2} \times \frac{N}{2}$ image. Thus, input for step 2 to 6 is an $\frac{N}{2} \times \frac{N}{2}$ image matrix. Similar to section 4.1, computation time of step 2 and step 4 is $O\left(\frac{N^2}{2}\right)$ and that of step 3 and 5 is $O\left(\frac{N^3}{4} + \frac{N}{2}\log_2\left(\frac{N}{2}\right)\right)$. Step 6 involves computation of inverse wavelet transform, which is an $O(N)$ function. Therefore, the computation complexity of encryption is:

$$T_{encryption} = O(N) + O\left(\frac{N^2}{2}\right) + O\left(\frac{N^3}{4} + \frac{N}{2}\log_2\left(\frac{N}{2}\right)\right) + O\left(\frac{N^2}{2}\right)$$

$$+ O\left(\frac{N^3}{4} + \frac{N}{2}\log_2\left(\frac{N}{2}\right)\right) + O(N)$$

$$= O\left(\frac{N^3}{2} + N^2 + N\log_2\left(\frac{N}{2}\right) + 2N\right)$$

$$\Rightarrow T_{encryption} = O\left(\frac{N^3}{2}\right)$$

**Decryption:** Image decryption process involves following steps:

1. Application of $DWT_2$ on the primary color components $g_i(x, y)$ of encrypted image.
2. Application of 2-D inverse FRT.
3. Decoding by conjugate of second CRPM.
4. Application of 2-D inverse FRT.
5. Decoding by conjugate of first CRPM.
6. Application of $IDWT_2$ on the R, G, and B channels obtained after step 5.

For an input image of size $N \times N$, step 1 is an $O(N)$ process which produces $\frac{N}{2} \times \frac{N}{2}$ image. Thus, input for step 2 to 6 is an $\frac{N}{2} \times \frac{N}{2}$ image matrix. Similar to section 4.1, computation time of step 2 and step 4 is $O\left(\frac{N^2}{2}\right)$ and that of step 3 and 5 is $O\left(\frac{N^3}{4} + \frac{N}{2}\log_2\left(\frac{N}{2}\right)\right)$. Step 6 involves computation of inverse wavelet transform, which is an $O(N)$ function. Therefore, the computation complexity of decryption is:

$$T_{deryption} = O(N) + O\left(\frac{N^2}{2}\right) + O\left(\frac{N^3}{4} + \frac{N}{2}\log_2\left(\frac{N}{2}\right)\right) + O\left(\frac{N^2}{2}\right)$$

$$+ O\left(\frac{N^3}{4} + \frac{N}{2}\log_2\left(\frac{N}{2}\right)\right) + O(N)$$

$$= O\left(\frac{N^3}{2} + N^2 + N\log_2\left(\frac{N}{2}\right) + 2N\right)$$

$$\Rightarrow T_{decryption} = O\left(\frac{N^3}{2}\right)$$

Thus, computation complexity of algorithm 4.3 is evaluated to:

$$T_{4.3} = T_{encryption} + T_{decryption} = O(N^3)$$

**4.4 Comparison with existing methods:**

For an image *I* of size *N* X *N*, as discussed in chapter 3, the computation complexity of Algorithms 3.1, 3.2 and 3.3 is $O(8N^3)$ i.e.

$$T_{3.1} \propto 8N^3,$$

$$T_{3.2} \propto 8N^3$$

$$\text{and } T_{3.3} \propto 8N^3$$

This chapter proposed three algorithms *viz.* algorithm 4.1, 4.2 and 4.3, in section 4.1, 4.2 and 4.3 respectively. Computation complexity of these algorithms is evaluated to be $O(N^3)$ i.e.

$$T_{4.1} \propto N^3,$$

$$T_{4.2} \propto N^3$$

$$\text{and } T_{4.3} \propto N^3$$

Therefore,

$$\frac{T_{4.1}}{T_{3.1}} \propto \left(\frac{N^3}{8N^3}\right) = \frac{1}{8}, \frac{T_{4.2}}{T_{3.2}} \propto \frac{1}{8}, \text{ and } \frac{T_{4.3}}{T_{3.3}} \propto \frac{1}{8}. \quad (4.4.1)$$

Based on equation (4.4.1), this work claims that the proposed algorithms 4.1, 4.2 and 4.3 are 8 time faster than the existing algorithms. This claim needs to be verified by conducting appropriate simulations. Following chapter discusses evidentory simulations in support to these calims.

# SIMULATION RESULTS

In order to investigate the quality of encryption, decryption and efficiency of proposed algorithms, digital simulations were performed in an environment as under:

**Processor –** Intel® Core™ i3 CPU, M380 2.53GHz.

**RAM –** 3.00 GB.

**Operating System –** 64-bit Windows 7.

**Simulation tool –** MATLAB® R2008b.

For investigation, two parameters were used to quantify the information required for comparison between existing algorithms and the proposed algorithms and these parameters are as under:

**(i) Mean Square Error:** Let $I = f(x, y)$ and $I' = f'(x, y)$ be the input image to and decrypted image from an algorithm respectively and let the size of $I$ and $I'$ be $M$x$N$. *MSE* between each of the primary color channels of $I$ and $I'$ is evaluated, which is formulated as follows [2]:

$$MSE_i = \frac{1}{M}\frac{1}{N}\sum_{x=1}^{M}\sum_{y=1}^{N}\left|f_i^{'}(x,y) - f_i(x,y)\right|^2 \qquad (5.1)$$

**(ii) Computation time:** Another parameter for evaluating the efficiency of algorithms is computation time. Chapter 3 & chapter 4 have analyzed various algorithms on the basis of their asymptotic computation complexity. The term computation time means total time required in encryption-decryption process of each algorithm for different fractional orders of *FRT*. On the basis of this collected data about computation time, a comparison between the existing and proposed algorithms is made in next sections.

The input image chosen for analysis is Lena (TIFF image; Size = 193 KB; pixel by pixel resolution = 256X256) as shown in fig. 5.1. Lena was encrypted using algorithm 4.1, 4.2 and 4.3 for $\alpha = \beta = \gamma = \delta = 0.5$ fractional orders of *FRT*. The resulting encrypted images obtained after execution of algorithm 4.1, 4.2 and 4.3 are shown in fig 5.2. These images can be decrypted on any fractional order of *FRT*, but the restored image may differ from the input image depending on the order of *FRT*. The decrypted images obtained using incorrect fractional orders of *FRT* are shown in fig. 5.3. Fig. 5.4 shows decrypted image obtained when correct fractional order of *FRT* are used.

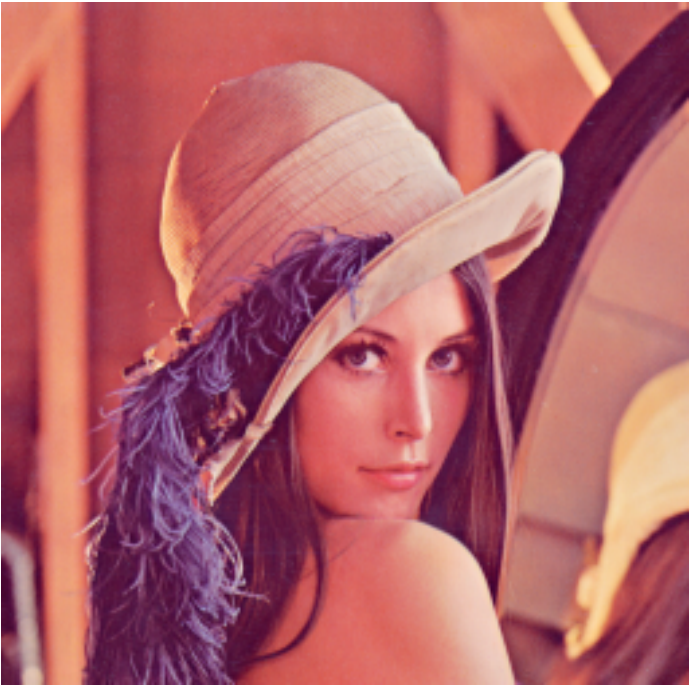

Fig 5.1. Input image to algorithm 3.1, 3.2 and 3.3 (Lena.tif, 256X256, color).

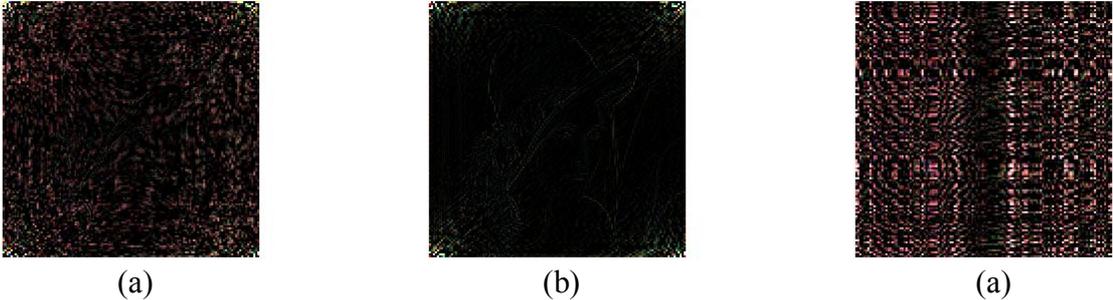

(a)            (b)            (a)

Fig 5.2. Encryted images of size 256X256 by (a) Algorithm 4.1 (b) Algorithm 4.2 and (c) Algorithm 4.3. The fractional orders of *FRT* are α = β = γ = δ = 0.5.

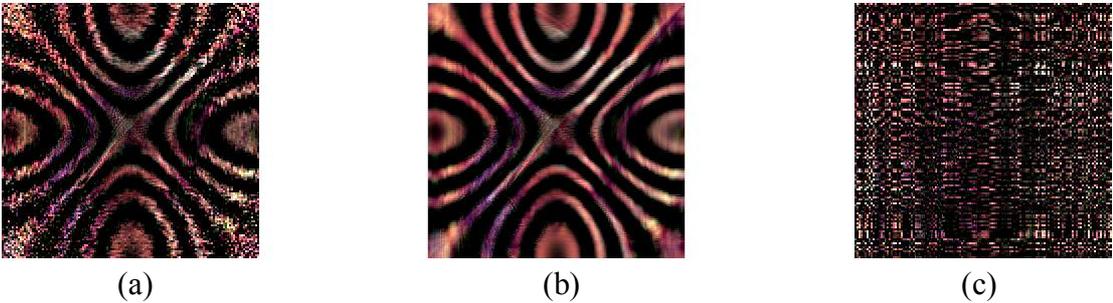

(a)            (b)            (c)

Fig 5.3. Decryted images of size 256X256 decrypted on an incorrect fractional order of inverse *FRT* by (a) Algorithm 4.1 (b) Algorithm 4.2 and (c) Algorithm 4.3. The fractional orders of *FRT* are α = β = γ = δ = 0.4.

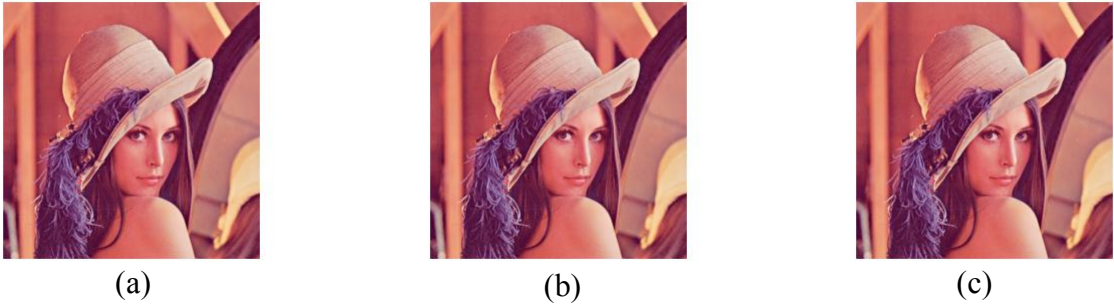

(a)      (b)      (c)

Fig 5.4. Decryted images of size 256X256 decrypted on appropriate fractional order of inverse *FRT* by (a) Algorithm 4.1 (b) Algorithm 4.2 and (c) Algorithm 4.3. The fractional orders of *FRT* are $\alpha = \beta = \gamma = \delta = 0.5$.

**TABLE 5.1**
**MSE BETWEEN DECRYPTED IMAGE AND THE INPUT IMAGE AS A FUNCTION OF ORDER OF *FRT*, USED FOR DECRYPTION IN ALGORITHM 3.1 AND ALGORITHM 4.1.**

| Order of *FRT* | MSE for Algorithm 3.1 (x$10^4$) | | | MSE for Algorithm 4.1 (x$10^4$) | | |
|---|---|---|---|---|---|---|
| | R MSE | G MSE | B MSE | R MSE | G MSE | B MSE |
| 0 | 6.94 | 2.50 | 2.43 | 6.79 | 2.43 | 2.37 |
| 0.1 | 6.94 | 2.50 | 2.43 | 6.88 | 2.46 | 2.40 |
| 0.2 | 6.92 | 2.49 | 2.42 | 6.89 | 2.47 | 2.40 |
| 0.3 | 6.92 | 2.49 | 2.42 | 6.85 | 2.44 | 2.39 |
| 0.4 | 6.82 | 2.45 | 2.39 | 6.61 | 2.39 | 2.34 |
| 0.5 | 2.68E-28 | 1.04E-28 | 1.02E-28 | 9.9E-29 | 4.5E-29 | 3.7E-29 |
| 0.6 | 6.82 | 2.45 | 2.39 | 6.67 | 2.41 | 2.36 |
| 0.7 | 6.93 | 2.49 | 2.43 | 6.78 | 2.42 | 2.37 |
| 0.8 | 6.92 | 2.49 | 2.42 | 6.91 | 2.47 | 2.41 |
| 0.9 | 6.90 | 2.49 | 2.41 | 6.92 | 2.47 | 2.41 |
| 1 | 6.94 | 2.49 | 2.43 | 6.81 | 2.43 | 2.37 |

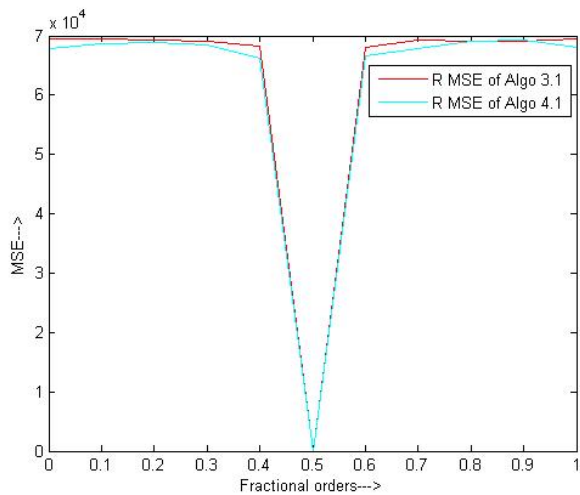

(a)

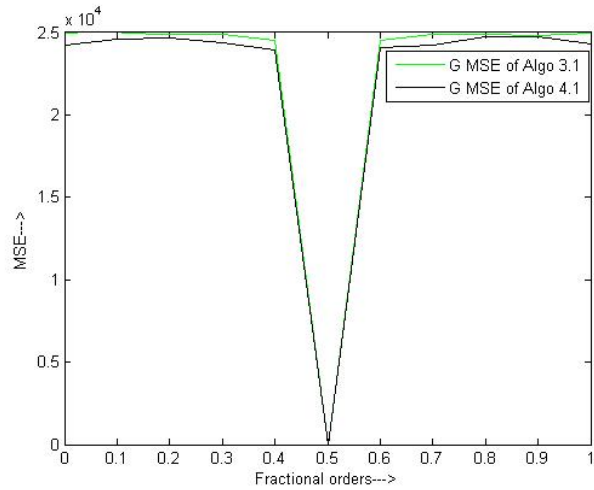

(b)

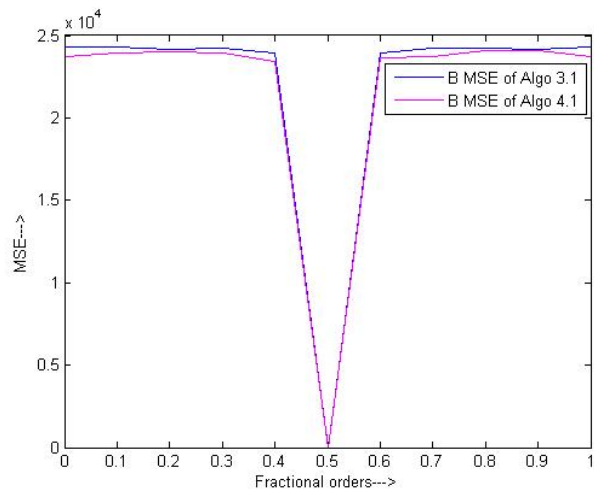

(c)

Fig 5.5. MSE graph showing the MSE between (a) R, (b) G and (c) B channels of decrypted image and the input image as a function of order of *FRT* used for decryption by using algorithm 3.1 and algorithm 4.1.

**TABLE 5.2**
**MSE BETWEEN DECRYPTED IMAGE AND THE INPUT IMAGE AS A FUNCTION OF ORDER OF *FRT* (4.0-6.0), USED FOR DECRYPTION IN ALGORITHM 4.1.**

| Order of *FRT* | MSE for Algorithm 4.1 (x$10^4$) | | | Order of *FRT* | MSE for Algorithm 4.1 (x$10^4$) | | |
|---|---|---|---|---|---|---|---|
| | R MSE | G MSE | B MSE | | R MSE | G MSE | B MSE |
| **4.0** | 6.64 | 2.41 | 2.35 | **5.1** | 2.97 | 1.05 | 0.97 |
| **4.1** | 6.84 | 2.47 | 2.40 | **5.2** | 6.65 | 2.39 | 2.26 |
| **4.2** | 6.69 | 2.37 | 2.33 | **5.3** | 6.54 | 2.38 | 2.33 |
| **4.3** | 6.36 | 2.23 | 2.22 | **5.4** | 6.01 | 2.13 | 2.14 |
| **4.4** | 6.73 | 2.33 | 2.30 | **5.5** | 6.71 | 2.32 | 2.29 |
| **4.5** | 6.72 | 2.31 | 2.29 | **5.6** | 6.71 | 2.33 | 2.30 |
| **4.6** | 6.00 | 2.14 | 2.14 | **5.7** | 6.36 | 2.24 | 2.22 |
| **4.7** | 6.54 | 2.38 | 2.34 | **5.8** | 6.71 | 2.37 | 2.33 |
| **4.8** | 6.68 | 2.39 | 2.26 | **5.9** | 6.84 | 2.46 | 2.39 |
| **4.9** | 2.98 | 1.05 | 0.98 | **6.0** | 6.61 | 2.4 | 2.34 |
| **5.0** | 9.9E-29 | 4.5E-29 | 3.7E-29 | | | | |

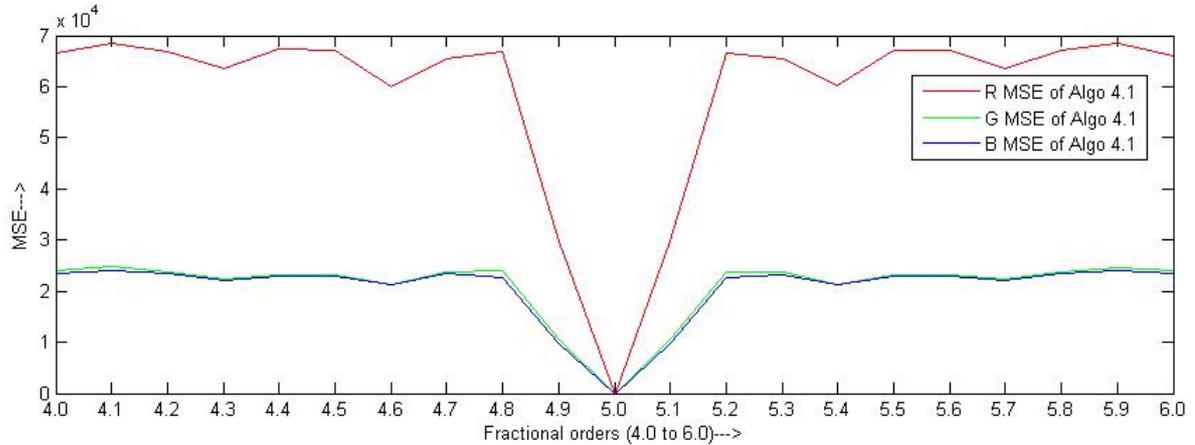

Fig 5.6. Variation in MSE between R, G and B channels of decrypted image and the input image as a function of order of *FRT* (4.0-6.0) used for decryption in algorithm 4.1. (See TABLE 5.2 for references).

**TABLE 5.3**
**MSE BETWEEN DECRYPTED IMAGE AND THE INPUT IMAGE AS A FUNCTION OF ORDER OF *FRT*, USED FOR DECRYPTION IN ALGORITHM 3.2 AND ALGORITHM 4.2.**

| Order of *FRT* | MSE for Algorithm 3.2 (X10$^4$) | | | MSE for Algorithm 4.2 (X10$^4$) | | |
|---|---|---|---|---|---|---|
| | R MSE | G MSE | B MSE | R MSE | G MSE | B MSE |
| 0 | 6.91 | 2.48 | 2.42 | 6.85 | 2.45 | 2.39 |
| 0.1 | 6.93 | 2.49 | 2.42 | 6.90 | 2.47 | 2.40 |
| 0.2 | 6.92 | 2.49 | 2.42 | 6.89 | 2.46 | 2.40 |
| 0.3 | 6.92 | 2.49 | 2.42 | 6.81 | 2.43 | 2.38 |
| 0.4 | 6.79 | 2.44 | 2.38 | 6.60 | 2.40 | 2.35 |
| 0.5 | 1.97E-28 | 7.98E-29 | 7.21E-29 | 9.38E-29 | 4.37E-29 | 3.31E-29 |
| 0.6 | 6.79 | 2.44 | 2.38 | 6.60 | 2.40 | 2.35 |
| 0.7 | 6.92 | 2.49 | 2.42 | 6.81 | 2.43 | 2.38 |
| 0.8 | 6.92 | 2.49 | 2.42 | 6.89 | 2.46 | 2.40 |
| 0.9 | 6.93 | 2.49 | 2.42 | 6.90 | 2.47 | 2.40 |
| 1 | 6.91 | 2.48 | 2.42 | 6.85 | 2.45 | 2.39 |

Fig. 5.5 shows a graph, comparing *MSE*'s generated by the algorithms 3.1 and 4.1 for different fractional orders of *FRT*. From the graph it is clear that both the algorithms perform equally in terms of the restoration of image. Similar results are observed for algorithm pair 3.2 - 4.2 and 3.3 - 4.3 and are shown in fig 5.7 and 5.9 respectively

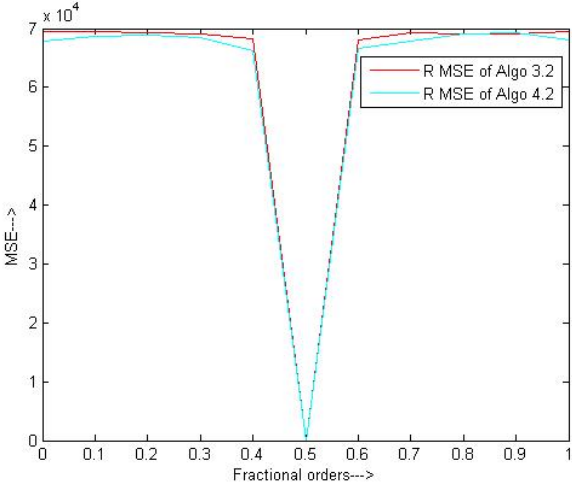

(a)

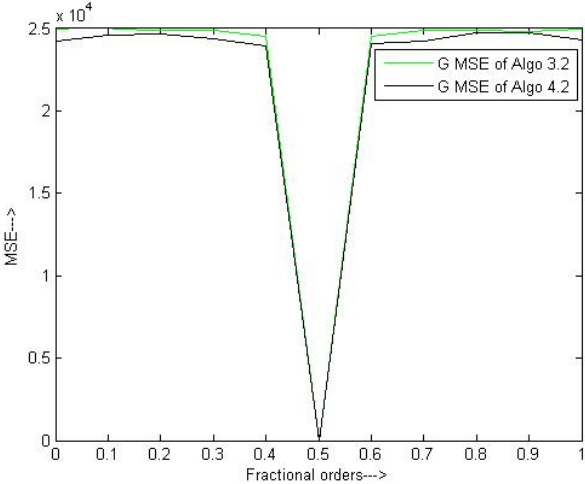

(b)

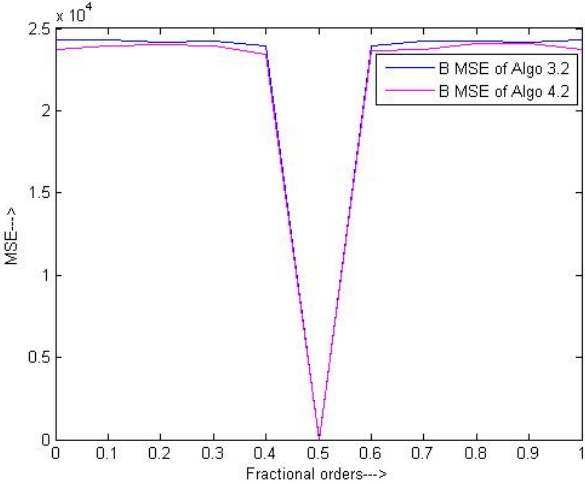

(c)

Fig 5.7. MSE graph showing the MSE between (a) R, (b) G and (c) B channels of decrypted image and the input image as a function of order of *FRT* used for decryption by using algorithm 3.2 and algorithm 4.2.

**TABLE 5.4**
**MSE BETWEEN DECRYPTED IMAGE AND THE INPUT IMAGE AS A FUNCTION OF ORDER OF *FRT* (4.0-6.0), USED FOR DECRYPTION IN ALGORITHM 4.2.**

| Order of *FRT* | MSE for Algorithm 4.2 ($\times 10^4$) | | | Order of *FRT* | MSE for Algorithm 4.2 ($\times 10^4$) | | |
|---|---|---|---|---|---|---|---|
| | R MSE | G MSE | B MSE | | R MSE | G MSE | B MSE |
| **4.0** | 6.60 | 2.40 | 2.35 | **5.1** | 2.99 | 1.06 | 0.98 |
| **4.1** | 6.86 | 2.48 | 2.41 | **5.2** | 6.75 | 2.42 | 2.28 |
| **4.2** | 6.65 | 2.35 | 2.32 | **5.3** | 6.48 | 2.37 | 2.33 |
| **4.3** | 6.23 | 2.19 | 2.18 | **5.4** | 5.79 | 2.06 | 2.07 |
| **4.4** | 6.69 | 2.31 | 2.28 | **5.5** | 6.65 | 2.27 | 2.25 |
| **.4.5** | 6.65 | 2.27 | 2.25 | **5.6** | 6.69 | 2.31 | 2.28 |
| **4.6** | 5.79 | 2.06 | 2.07 | **5.7** | 6.23 | 2.19 | 2.18 |
| **4.7** | 6.48 | 2.37 | 2.33 | **5.8** | 6.65 | 2.35 | 2.32 |
| **4.8** | 6.75 | 2.42 | 2.28 | **5.9** | 6.86 | 2.48 | 2.41 |
| **4.9** | 2.99 | 1.06 | 0.98 | **6.0** | 6.60 | 2.40 | 2.35 |
| **5.0** | 9.38E-29 | 4.37E-29 | 3.31E-29 | | | | |

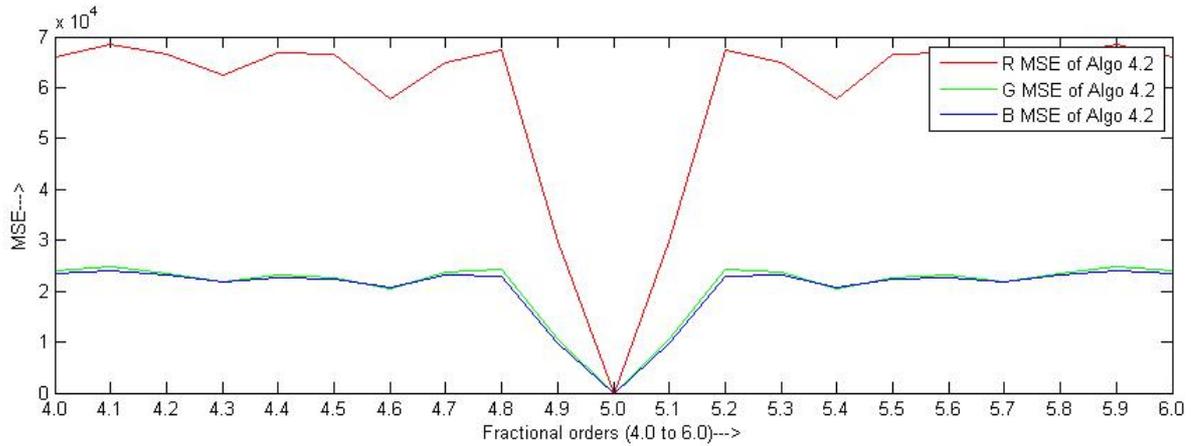

Fig 5.8. Variation in MSE between R, G and B channels of decrypted image and the input image as a function of order of *FRT* (4.0-6.0) used for decryption in algorithm 4.2. (See TABLE 5.4 for references).

.Fig. 5.11, 5.12 and 5.13 consist of graphs comparing computation time required by algorithm pair 3.1 Vs 4.1, 3.2 Vs 4.2 and 3.3 Vs 4.3. From these graphs based on the tables 5.4, 5.5 and 5.6 respectively, it is observed that the ratio between computation times of algorithms 4.1 and 3.1 is ≈1:8. Similar results are observed between algorithm pairs 3.2 - 4.2 and 3.3 - 4.3.

**TABLE 5.5**

**MSE BETWEEN DECRYPTED IMAGE AND THE INPUT IMAGE AS A FUNCTION OF ORDER OF *FRT*, USED FOR DECRYPTION IN ALGORITHM 3.3 AND ALGORITHM 4.3.**

| Order of *FRT* | MSE for Algorithm 3.3 (X$10^4$) | | | MSE for Algorithm 4.3 (X$10^4$) | | |
|---|---|---|---|---|---|---|
| | R MSE | G MSE | B MSE | R MSE | G MSE | B MSE |
| 0 | 6.93 | 2.49 | 2.43 | 1.71 | 0.51 | 0.52 |
| 0.1 | 6.96 | 2.49 | 2.43 | 1.76 | 0.52 | 0.53 |
| 0.2 | 6.93 | 2.49 | 2.43 | 1.89 | 0.56 | 0.58 |
| 0.3 | 6.92 | 2.48 | 2.42 | 1.93 | 0.57 | 0.57 |
| 0.4 | 6.95 | 2.50 | 2.43 | 2.00 | 0.59 | 0.59 |
| 0.5 | 2.94E-28 | 1.06E-28 | 1.04E-28 | 5.22E-29 | 1.6E-29 | 1.6E-29 |
| 0.6 | 6.97 | 2.50 | 2.44 | 1.97 | 0.59 | 0.58 |
| 0.7 | 6.91 | 2.48 | 2.42 | 2.07 | 0.63 | 0.62 |
| 0.8 | 6.92 | 2.49 | 2.42 | 1.74 | 0.54 | 0.53 |
| 0.9 | 6.97 | 2.50 | 2.44 | 1.86 | 0.58 | 0.58 |
| 1 | 6.95 | 2.50 | 2.43 | 1.74 | 0.54 | 0.54 |

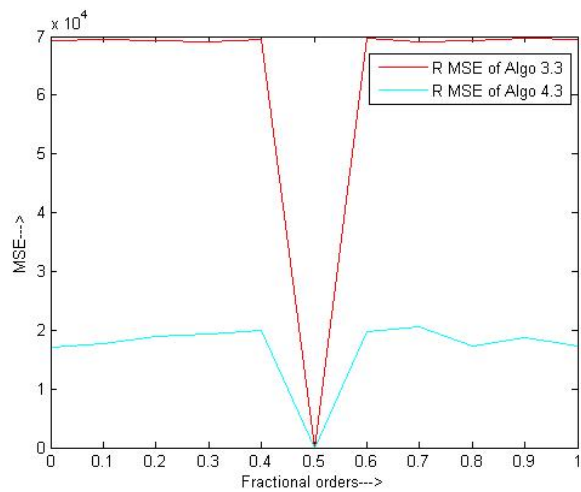

(a)

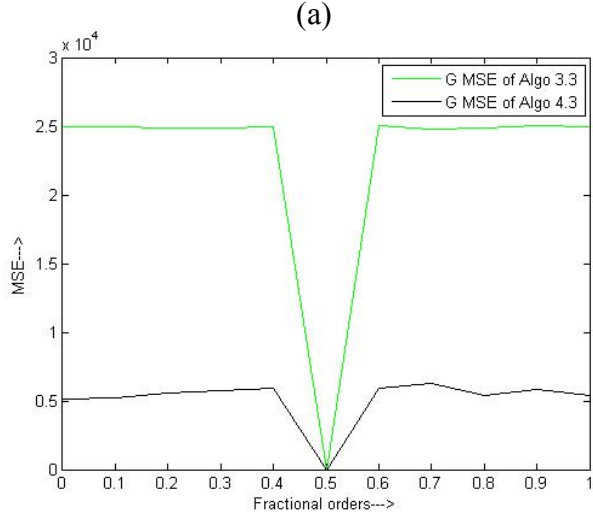

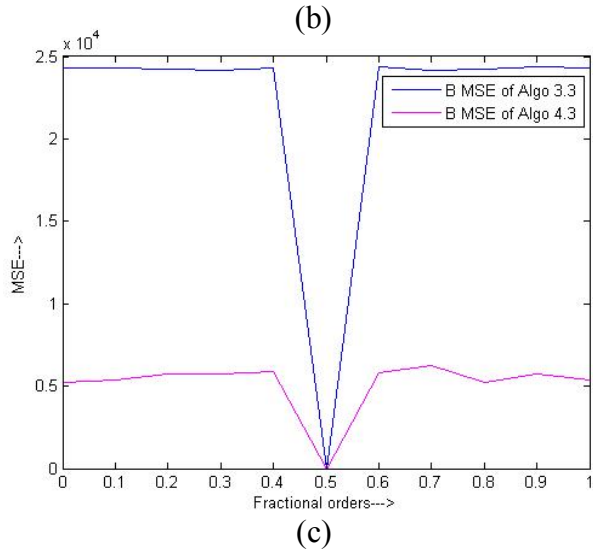

(b)

(c)

Fig 5.9. MSE graph showing the MSE between (a) R, (b) G and (c) B channels of decrypted image and the input image as a function of order of *FRT* used for decryption by using algorithm 3.3 and algorithm 4.3.

**TABLE 5.6**
**MSE BETWEEN DECRYPTED IMAGE AND THE INPUT IMAGE AS A FUNCTION OF ORDER OF *FRT* (4.0-6.0), USED FOR DECRYPTION IN ALGORITHM 4.3.**

| Order of *FRT* | MSE for Algorithm 4.3 ($\times 10^4$) | | | Order of *FRT* | MSE for Algorithm 4.3 ($\times 10^4$) | | |
|---|---|---|---|---|---|---|---|
| | R MSE | G MSE | B MSE | | R MSE | G MSE | B MSE |
| **4.0** | 2.00 | 0.59 | 0.59 | **5.1** | 1.25 | 0.34 | 0.33 |
| **4.1** | 1.92 | 0.57 | 0.56 | **5.2** | 1.74 | 0.51 | 0.49 |
| **4.2** | 1.84 | 0.55 | 0.53 | **5.3** | 1.86 | 0.56 | 0.54 |
| **4.3** | 1.80 | 0.52 | 0.52 | **5.4** | 1.96 | 0.59 | 0.57 |
| **4.4** | 1.84 | 0.54 | 0.52 | **5.5** | 1.97 | 0.59 | 0.57 |
| **.4.5** | 1.92 | 0.56 | 0.56 | **5.6** | 1.93 | 0.57 | 0.56 |
| **4.6** | 1.96 | 0.58 | 0.57 | **5.7** | 1.82 | 0.54 | 0.52 |
| **4.7** | 1.88 | 0.55 | 0.54 | **5.8** | 1.80 | 0.54 | 0.52 |
| **4.8** | 1.72 | 0.50 | 0.49 | **5.9** | 1.89 | 0.57 | 0.55 |
| **4.9** | 1.25 | 0.34 | 0.33 | **6.0** | 1.97 | 0.59 | 0.58 |
| **5.0** | 5.22E-29 | 1.6E-29 | 1.6E-29 | | | | |

**TABLE 5.7**
**COMPUTATION TIME (IN SECONDS) OF ALGORITHM 3.1 AND ALGORITHM 4.1.**

| Order of *FRT* | | 0 | 0.1 | 0.2 | 0.3 | 0.4 | 0.5 | 0.6 | 0.7 | 0.8 | 0.9 | 1.0 |
|---|---|---|---|---|---|---|---|---|---|---|---|---|
| Computation Time (in Sec.) | Algo 3.1 | 72.3 | 73.1 | 73.1 | 72.9 | 73.0 | 72.8 | 72.8 | 72.9 | 72.9 | 72.8 | 72.9 |
| | Algo 4.1 | 9.3 | 9.3 | 9.2 | 9.2 | 9.2 | 9.2 | 9.3 | 9.2 | 9.3 | 9.2 | 9.2 |
| **Ratio** | T3.1/T4.1 | 7.8 | 7.9 | 7.9 | 7.9 | 7.9 | 7.9 | 7.9 | 7.9 | 7.9 | 7.9 | 7.9 |

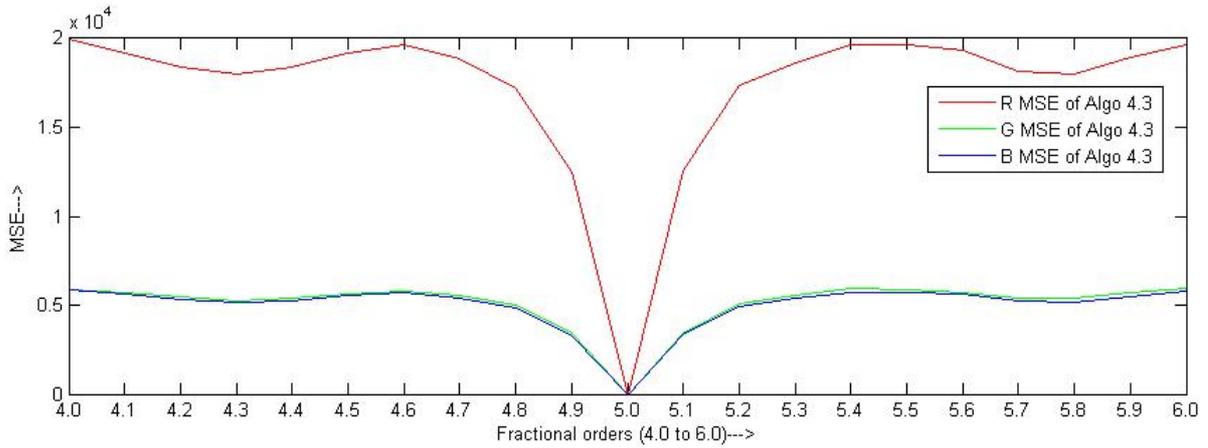

Fig 5.10. Variation in MSE between R, G and B channels of decrypted image and the input image as a function of order of *FRT* (4.0-6.0) used for decryption in algorithm 4.3. (See TABLE 5.6 for references).

**TABLE 5.8**
**COMPUTATION TIME (IN SECONDS) OF ALGORITHM 3.2 AND ALGORITHM 4.2.**

| Order of *FRT* | | 0 | 0.1 | 0.2 | 0.3 | 0.4 | 0.5 | 0.6 | 0.7 | 0.8 | 0.9 | 1.0 |
|---|---|---|---|---|---|---|---|---|---|---|---|---|
| Computation Time (in Sec.) | Algo 3.2 | 73.4 | 73.3 | 73.2 | 73.2 | 73.4 | 73.4 | 73.2 | 73.2 | 73.2 | 73.4 | 73.2 |
| | Algo 4.2 | 9.4 | 9.3 | 9.3 | 9.3 | 9.3 | 8.8 | 9.3 | 9.3 | 9.3 | 9.3 | 9.3 |
| Ratio | T3.2/T4.2 | 7.8 | 7.9 | 7.9 | 7.9 | 7.9 | 8.3 | 7.9 | 7.9 | 7.9 | 7.9 | 7.9 |

**TABLE 5.9**
**COMPUTATION TIME (IN SECONDS) OF ALGORITHM 3.3 AND ALGORITHM 4.3.**

| Order of *FRT* | | 0 | 0.1 | 0.2 | 0.3 | 0.4 | 0.5 | 0.6 | 0.7 | 0.8 | 0.9 | 1.0 |
|---|---|---|---|---|---|---|---|---|---|---|---|---|
| Computation Time (in Sec.) | Algo 3.3 | 72.6 | 73.0 | 72.9 | 73.1 | 72.9 | 72.9 | 72.9 | 73.0 | 72.9 | 72.9 | 73.0 |
| | Algo 4.3 | 9.2 | 9.2 | 9.2 | 9.2 | 9.2 | 9.2 | 9.2 | 9.2 | 9.2 | 9.2 | 9.2 |
| Ratio | T3.2/T4.2 | 7.9 | 7.9 | 7.9 | 7.9 | 7.9 | 7.9 | 7.9 | 7.9 | 7.9 | 7.9 | 7.9 |

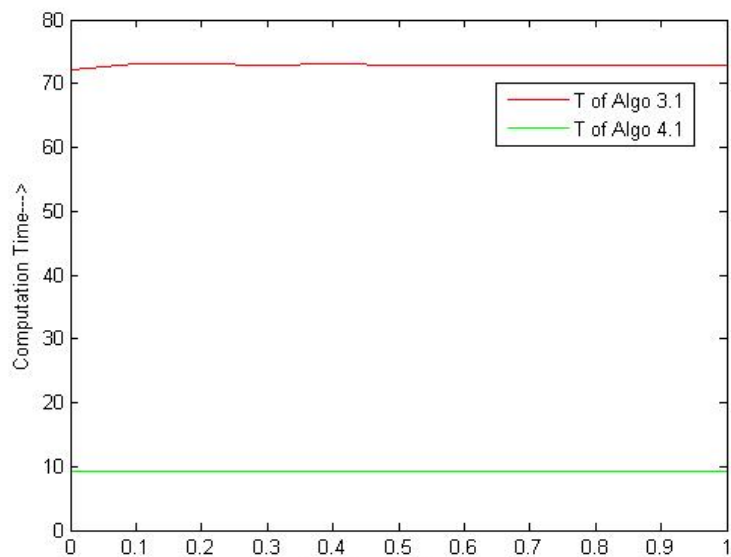

Fig 5.11. Graph showing computation time of algorithm 3.1 and algorithm 4.1 as a function of order of *FRT*.

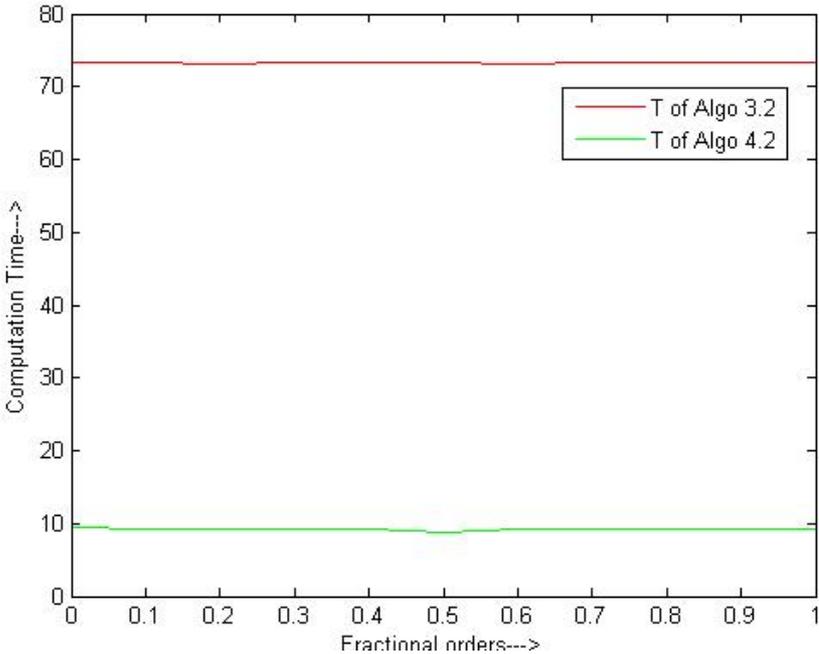

Fig 5.12. Graph showing computation time of algorithm 3.2 and algorithm 4.2 as a function of order of *FRT*.

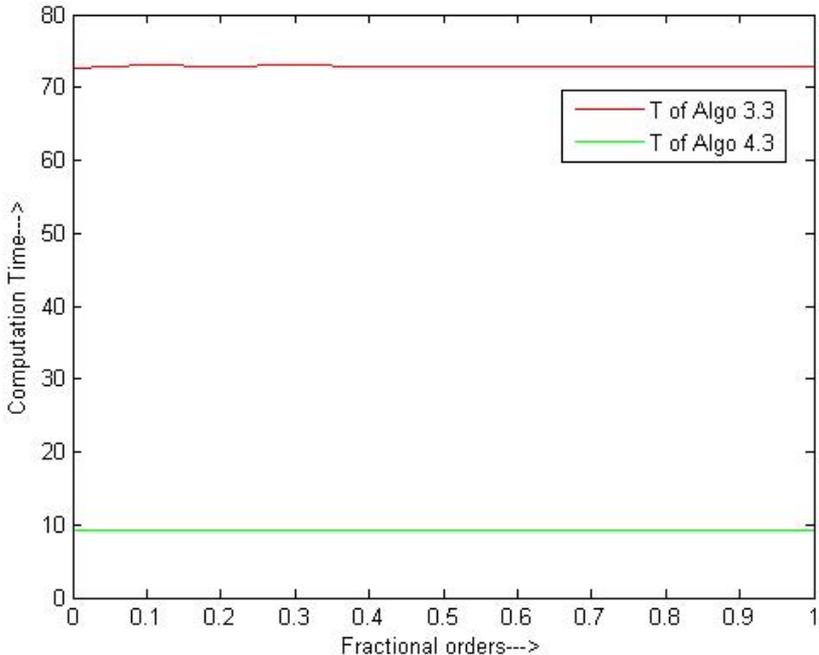

Fig 5.13. Graph showing computation time of algorithm 3.3 and algorithm 4.3 as a function of order of *FRT*.

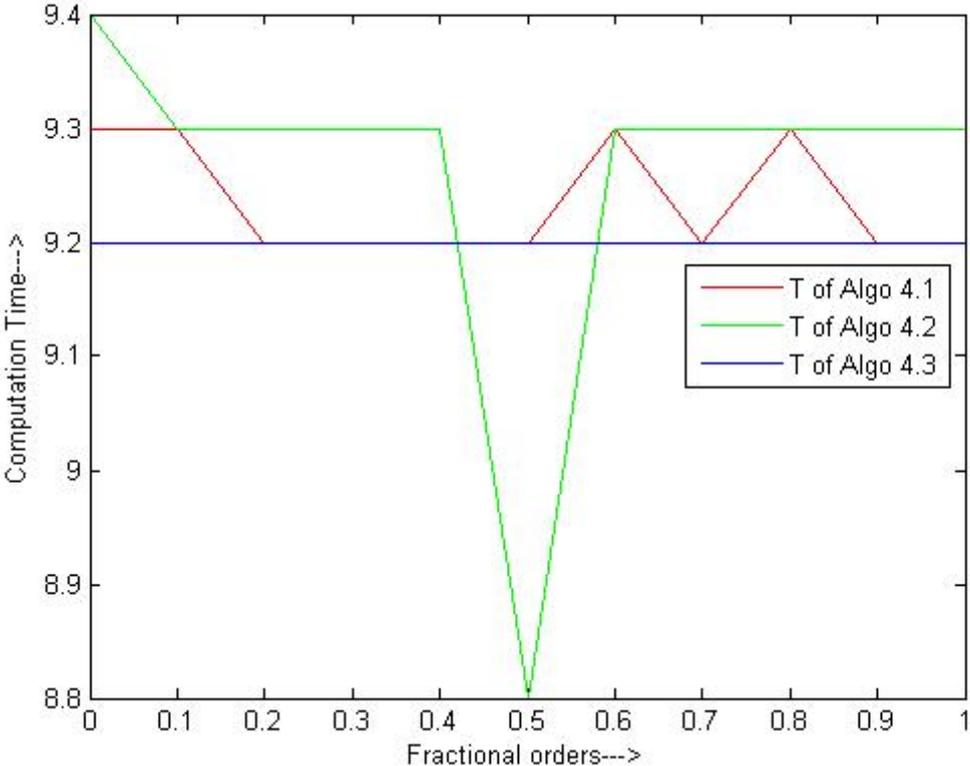

Fig 5.14. Graph showing computation time (in seconds) of algorithm 4.1, 4.2 and 4.3 as a function of order of *FRT*.

Fig 5.14 shows a graph, comparing computation time required by algorithm 4.1, 4.2 and 4.3 for different fractional orders of *FRT*. From this figure, it is shown that the computation time required by Algorihtm 4.1, 4.2 and 4.3 is nearly equal (ranging between 8.8 second to 9.4 second). The variation in computation time may be caused by other background processes running on machine.

# CONCLUSION

This work proposed three algorithms *viz*. Algorithm 4.1, 4.2 and 4.3 in section 4.1, 4.2 and 4.3 respectively for image encryption and decryption based on fractional Fourier transform (*FRT*), discrete Wavelet transform and Chaos functions. *MSE* and computation time are taken as the parameters for performance-comparison among these algorithms Section 4.4 claimed in equation 4.4.1 that these algorithms 4.1, 4.2 and 4.3 should be nearly 8 times faster than algorithms 3.1, 3.2 and 3.3 respectively. This improvement in computation time is due to reduced input data size (or image size) for encryption and decryption by a factor of 4. For this purpose image compression characteristic of the discrete wavelet transform are utilized. This claim has been verified in this work.

For verification of claims made in chapter 4, simulations were run using *MATLAB*7.7. Results obtained after these simulations were summarized in chapter 5. Table 5.1, 5.3 and 5.5 summarize the *MSE* between R, G and B channels of input and restored image for different fractional orders of *FRT*, using algorithm pair 3.1-4.1, 3.2-4.2 and 3.3-4.3. From these tables the *MSE* for algorithm 4.1, 4.2 and 4.3 is found to be consistently less than their counterparts, hence the proposed algorithms are less likely to information-loss during encryption-decryption process.

Table 5.2, 5.4 and 5.6 present a deeper insight into the variation of *MSE* with respect to the fractional orders of *FRT* (ranging between 4.0 and 6.0 and at an interval of $1 \times 10^{-1}$). These results are shown by graphs of fig.5.6, 5.8 and 5.10 respectively. These graphs indicate that the correct decryption of image is possible within 0.49 to 0.51 value of α, β, γ and δ as compared to earlier range of 0.45 to 0.55 reported earlier

**Future work:**

a. Faster algorithms for computation of 2-*D FRT*, 2-*D* inverse *FRT*, $DWT_2$ and $IDWT_2$ may result in further reduction of computation.

b. This work can be extended for different formats of images.

c. This work may be extended using other transforms methods also.

## PUBLICATIONS

[1] Prerana Sharma, Devesh Mishra, Ankur Agarwal, "Efficient Image Encryption and Decryption Using Discrete Wavelet Transform and Fractional Fourier Transform", *Fifth ACM International Conference on Security of Information and Networks 2012 SIN2012,* pp 153-157, 2012.